\documentclass[fleqn,usenatbib,useAMS]{mnras}

\usepackage{graphicx}	
\usepackage{amsmath}	
\usepackage{amssymb}	
\usepackage{multicol}        
\usepackage{bm}		
\usepackage{pdflscape}	
\usepackage{float}
\usepackage{amsfonts}
\usepackage{booktabs}
\usepackage{siunitx}

\usepackage[T1]{fontenc}
\usepackage{ae,aecompl}

\usepackage{newtxtext,newtxmath}

\title[Single Fluid vs. Multifluid]{Single Fluid vs. Multifluid: Comparison between single fluid and multifluid dust models for disc planet interactions}

\author[K . Chan and S-J . Paardekooper]{Kevin Chan$^{1}$ and Sijme-Jan Paardekooper$^{1,2}$\thanks{Contact e-mail: \href{mailto:k.chan@qmul.ac.uk}{k.chan@qmul.ac.uk}}
\\
$^{1}$Astronomy Unit, School of Physics and Astronomy,Queen Mary University of London, Bethnal Green, London E1 4NS\\
$^{2}$DAMTP, University of Cambridge, Wilberforce Road, Cambridge CB3 0WA, UK}

\date{Last updated 2022 February 22; in original form N/A}

\pubyear{2022}

\begin{document}
\label{firstpage}
\pagerange{\pageref{firstpage}--\pageref{lastpage}}
\maketitle

\begin{abstract}
Recent observations of substructures such as dust gaps and dust rings in protoplanetary discs have highlighted the importance of including dust into purely gaseous disc models. At the same time, computational difficulties arise with the standard models of simulating the dust and gas separately. These include the cost of accurately simulating the interactions between well coupled dust and gas and issues of dust concentration in areas below resolution of the gas phase. We test a single fluid approach, that incorporates the terminal velocity approximation valid for small particles, which can overcome these difficulties, through modification of FARGO3D.
We compare this single fluid model with a multifluid model for a variety of planet masses. We find differences in the dust density distribution in all cases. For high-mass, gap-opening planets we find differences in the amplitude of the resulting dust rings, which we attribute to the failure of the terminal velocity approximation around shocks. For low-mass planets, both models agree everywhere except in the corotation region, where the terminal velocity approximation shows overdense dust lobes. We tentatively interpret these as dusty equivalents of thermal lobes seen in non-isothermal simulations with thermal diffusion, but more work is necessary to confirm this. At the same resolution, the computational time for the terminal velocity approximation model is significantly less than a two fluid model. We conclude that the terminal velocity approximation is a valuable tool for modelling a protoplanetary disc but that care should be taken when shocks are involved.
\end{abstract}




\begingroup
\let\clearpage\relax

\endgroup
\newpage

\section{Introduction}

In order to simulate the evolution of protoplanetary discs and its interaction with embedded planets accurately, many physical effects have been researched and implemented into hydrodynamic disc simulations over the years. These works explore the effects of radiation and the energy budget \citep[e.g.][]{2006A&A...459L..17P, 2008A&A...487L...9K, lega2014, 2017MNRAS.472.4204M}, magnetic turbulence (e.g. \citealt{nelson03}; \citealt{flock17}), presence of an ordered magnetic field (e.g. \citealt{baruteau11}; \citealt{mcnally17}) and dynamical torques on an embedded planet (e.g. \citealt{paadekooper14}). At the same time, most studies have been based on the simplification of a pure gaseous disc due to the low dust to gas ratio where the dust canonically only comprises $1\%$ of the disc itself (\citealt{bohlin78}). This approximation has been effective in modelling planet disc interactions and improving our understanding of important factors such as migration of planets (\citealt{masset08}). More recently however, dust has been included when simulating protoplanetary discs, mainly to compare with the observations of sub-structures by \cite{alma15}. The most well-known observations of discs are around the young star HL Tau and TW Hydrae (\citealt{andrews16}) where dust rings and gaps have been observed by ALMA with questions over the presence of planets within the gaps itself. Previous studies have implemented dust, evolving separately from the gas, as a cold pressureless fluid (\citealt{paardekooper04}; \citealt{rodenkirch21}; \citealt{drazkowska19}) or as separate particles (\citealt{mcnally19b}; \citealt{baruteau21}; \citealt{rowther20}) which interact with the gas phase through a drag force. The drag force is for example responsible for the effect of radial drift of dust towards pressure maxima (\citealt{whipple72}). 

The inclusion of dust and the effects of dust radially drifting have piqued theories of planetesimals forming through dynamical instabilities in pressure bumps (e.g. \citealt{johansen09}). However, the dust back reaction on the gas has often been ignored, in particular in simulations with embedded planets \citep{paardekooper04, mcnally19b}. Over the years, studies have shown the importance of including the dust and the back reaction (e.g. \citealt{kanagawa17}; \citealt{dipierro18a}; \citealt{gonzalez18}). They show that even with a small dust to gas ratio, the viscous evolution of a gaseous disc can change significantly due to the radial drift of both the gas and the dust. The dynamics and the evolution of a protoplanetary disc in the inner and outer regions are shown to both be affected differently due to the presence of dust. The simulations of \cite{kanagawa17} show that the dust feedback on the gas slows the radial velocity of the gas as it travels inwards. At low viscosity or high dust to gas ratios, the gas can flow outwards and may cause the dust to gas mass ratio in the inner regions of the disc to exceed unity. \cite{dipierro18a} investigate steady-state dynamics of a protoplanetary disc with multiple species of dust rather than a single population. Their results differ from \cite{kanagawa17} in that for small dust to gas ratios the gas outflow can still occur in the outer regions due to the cumulative feedback from the dust species. In the inner regions, the dust inward motion is reduced below the gas inward motion due to strong drag and dust settling. In both cases we see that the dust back reaction on the gas is important to accurately evolve a disc that contains both dust and gas.

 Since the first results from \cite{alma15} of the HL Tau region, more protoplanetary discs with substructures of dust rings have been observed. This includes observations in the Taurus star forming region by \cite{long18} and other regions through the Disk Substructures at High Angular Resolution Project (DSHARP) conducted with ALMA \citep{andrews18}, which provides high resolution data of 20 protoplanetary discs. From these observations there have been works to link the dust rings in these regions to the presence of planets \citep[see e.g.][]{zhang18}. Additionally there has been work on trying to identify signatures of a migrating planet by the properties of the rings itself (e.g. \citealt{meru18}; \citealt{nazari20}), and the possibility of a second gap made from the same planet was explored in \cite{gonzalez15}. 
 
 The recent observations and studies are driving more and more hydrodynamic disc simulations to include dust. 
 This has been done both for Lagrangian smoothed-particle hydrodynamics, SPH (\citealt{laibe12}; \citealt{ayliffe12}), and Eulerian grid-based methods, where quantities such as the velocity and density are stored at fixed points in the fluid as it passes by (\citealt{llambay16}). For a comparisons between the two methods see  \cite{price10}. In the Eulerian formalism, each dust size is simulated as a separate pressureless fluid \citep{llambay19}. When considering only a single dust size, we are in the two-fluid regime (gas and dust). 
 
 However, several disadvantages have been identified for the widely used two fluid formalism. One particular disadvantage is that dust particles can concentrate in areas below the resolution length of the gas phase (\citealt{price10}; \citealt{ayliffe12}). When this occurs they no longer feel the differential forces from the gas, being held there indefinitely. In addition to this, the two fluid models are expensive to run as we have to take into account the interactions between the two fluids such as the drag. These interactions, in particular for small, well-coupled particles, happen on a small spatial scale and short timescales. In order to model accurately the relative velocities, we require a high spatial resolution in order to simulate the dissipation rate between the gas and dust correctly \citep{laibe14}. Lastly, we require infinitely small time steps, when the drag time scale tends tends to zero in the limit of perfect coupling, to maintain stability in explicit methods for numerical modelling as at this point, the dust and gas move as a single fluid. Overall this creates a model which can be taxing in terms of computational cost.

These disadvantages were the motivations for a simplified view of the two fluid formalism into a one fluid model by \cite{laibe14}. This is through taking the two fluid mixture of dust and gas, and describing it as a single fluid moving with the barycentric velocity of the two phases. An extension of this is under the strong drag regime whereby the dust is tightly coupled with the gas, "terminal velocity approximation" (\citealt{youdin05}, \citealt{jacquet11}). 

The terminal velocity approximation (TVA) has proven to be very useful when considering tightly coupled gas and dust in numerical simulations providing a clear advantage in the strong drag regime compared to two-fluid models. The TVA model has been implemented in SPH code (\citealt{ballabio18}; \citealt{chen18}; \citealt{vericel21}) and in grid-based code (\citealt{lovascio19}) with many studies reproducing ALMA observations with varying planet sizes embedded for discs around young stars. These studies include HL Tau (\citealt{dipierro15}), Elias 24 (\citealt{dipierro18b}), CQ Tau (\citealt{gabellini19}), DS Tau (\citealt{veronesi20}), HD143006 (\citealt{ballabio21}), HD100546 (\citealt{fedele21}) and PDS 70 (\citealt{toci20}). The TVA model has been used extensively in a large variety of studies which has proven the usefulness of the model in many other situations. Examples of other areas include the role of dust radial drift (\citealt{toci21}), dust entrainment by photoevaporative winds (\citealt{hutchison16}), dusty vortices (\citealt{lovascio19}), polydisperse streaming instability (\citealt{paardekooper20}), magnetohydrodynamic methods for dust-gas mixture (\citealt{tsukamoto21}), constraining disc to star mass ratio (\citealt{terry22}), flybys in protoplanetary discs (\citealt{cuello19}; \citealt{cuello20}) and transitional discs (\citealt{ragusa17}). 

\cite{lin17} adapted the TVA model through showing that the evolution of the dust ratio can be recast as an effective energy equation when the gas follows a locally isothermal equation of state. This is when the disc is kept at a fixed temperature based on the position due to a constant sound speed profile. The reasoning behind this approximation is that under  strong  drag  and  isothermal  gas,  the  dust is transported  through  the  flow  of  the  gas  which  is similar to the advection of entropy in an adiabatic fluid (\citealt{lin17}). In addition we can see that the dust and gas drag  causes  exchange  of  dust  density  of  a  fluid parcel with its surroundings. This is similar to heat exchange between the gas parcel and it’s surroundings  as  the  dust  is  treated  as  a  cold  pressureless fluid, providing an effective heating/cooling effect. Therefore,  in  a  way  the  isothermal  gas  behaves adiabatically  with  the  evolution of the dust ratio being similar to an energy evolution equation. 

Implementation of this model has been included in the study of low mass planet and disc interactions (\citealt{chen18}) and dusty vortices (\citealt{lovascio19}). These studies have been important in showing how the model can be implemented into a hydrocode. At the same time, \cite{lovascio19} has found that the one fluid, terminal velocity approximation model breaks down around shocks. Given the advantages and wide use of the TVA for modelling embedded planets, it is our aim to test the limits of the locally isothermal terminal velocity approximation when simulating the evolution of a dust and gas mixture around a large planet where shocks are most prominent due to the tidal interactions between the planet and the disc. 

In this paper we are extending the \cite{lovascio19} modification of the hydrocode FARGO3D (\citealt{llambay16}) by implementing a locally isothermal equation of state for the terminal velocity approximation and the ability to evolve the dust and gas on a global disc in cylindrical coordinates. Through this, we will be comparing our modification to the full two fluid model using FARGO Multifluid \citep{llambay19}. The main comparisons will be the computational cost, evolution of the dust for different dust sizes modified through the constant Stokes number used in FARGO, different dust to gas ratios and planet sizes embedded in the disc. The paper is structured as follows. In section \ref{FARGO}, we show the evolution equations for the multifluid model and the equations for single fluid locally isothermal terminal velocity approximation (LITVA), taking the barycentric values of the well coupled species. Section \ref{method} refers to the implementation and simulation setup for both multifluid and single fluid models. In section \ref{zero}, we present the results comparing LITVA to FARGO Multifluid and show the limitations of the terminal velocity approximation. In section \ref{discuss}, we discuss our results and the future of this model and present our conclusions in section \ref{conclusions}.

\section{Multifluid and Single Fluid Models} \label{FARGO}

\subsection{FARGO Multifluid}

Starting off with the FARGO3D Multifluid implementation (\citealt{llambay19}), we are considering just one species of dust and gas in a two fluid system. The equations to be solved in the locally isothermal regime are the continuity equations for the gas and dust respectively,

\begin{equation} \label{gascont}
    \frac{\partial \rho_{\rm g}}{\partial t} + \nabla \cdot (\rho_{\rm g} \textbf{u}_{\rm g}) = 0,
\end{equation}

\begin{equation}
    \frac{\partial \rho_{\rm d}}{\partial t} + \nabla \cdot (\rho_{\rm d} \textbf{u}_{\rm d}) = 0,
\end{equation}

\noindent and the momentum equations,

\begin{equation}
    \frac{\partial (\rho_{\rm g} \textbf{u}_{\rm g})}{\partial t} + \nabla \cdot (\rho_{\rm g} \textbf{u}_{\rm g} \textbf{u}_{\rm g}) =  K(\textbf{u}_{\rm d}-\textbf{u}_{\rm g}) - \nabla P_{\rm g} + \rho_{\rm g} \textbf{f},
\end{equation}

\begin{equation} \label{dustmom}
    \frac{\partial (\rho_{\rm d} \textbf{u}_{\rm d})}{\partial t} + \nabla \cdot (\rho_{\rm d} \textbf{u}_{\rm d}\textbf{u}_{\rm d}) =  - K(\textbf{u}_{\rm d}-\textbf{u}_{\rm g}) + \rho_{\rm d} \textbf{f}, 
\end{equation}

\noindent with the subscripts ${\rm d}$ and ${\rm g}$  referring to the dust and the gas respectively, $K$ being the drag coefficient, the pressure is $P_{\rm g}=c_{\rm s}^2 \rho_{\rm g}$, with $c_{\rm s}$ being the gas sound speed and a general body force $\textbf{f}$ such as gravity. In the implementation of FARGO3D, these equations are solved through the method of operator splitting (\citealt{stone92}). 

\subsection{Single Fluid Model}

The two fluid equations (\ref{gascont})-(\ref{dustmom}) can be reformulated into a single fluid. The next sections follow the same process as \cite{laibe14} where the gas and dust are moving with a barycentric velocity and a relative velocity is defined between the two couple fluids,

\begin{equation} \label{relvel}
    \textbf{u} = \frac{\rho_{\rm g} \textbf{u}_{\rm g} + \rho_{\rm d} \textbf{u}_{\rm d}}{\rho_{\rm g} +\rho_{\rm d}},
\end{equation}

\begin{equation}
    \Delta\textbf{u} \equiv \textbf{u}_{\rm d} - \textbf{u}_{\rm g},
\end{equation}
Since we are dealing with the barycentric frame, we will be considering the total density of the mixture given simply as $\rho=\rho_{\rm g}+\rho_{\rm d}$. Another important parameter is the evolution of the dust to gas ratio, $\rho_{\rm d}/\rho_{\rm g}$.

Rearranging equation (\ref{relvel}), we obtain the identities,

\begin{equation} \label{gasvel}
    \textbf{u}_{\rm g} = \textbf{u} - \frac{\rho_{\rm d}}{\rho} \Delta\textbf{u},
\end{equation}

\begin{equation}\label{dustvel}
    \textbf{u}_{\rm d} = \textbf{u} + \frac{\rho_{\rm g}}{\rho}\Delta\textbf{u},
\end{equation}
In addition to (\ref{gasvel}) and (\ref{dustvel}), we can simplify the two fluid equations through the use of the dust fraction $f_{\rm d}$, the gas fraction $\rho_{\rm g}/\rho$ and the stopping time $t_{s}$, which is the decay timescale for the relative velocity between dust and gas,

\begin{equation}\label{dustfrac}
    f_{\rm d} = \frac{\rho_{\rm d}}{\rho},
\end{equation}

\begin{equation}\label{gasfrac}
    \frac{\rho_{\rm g}}{\rho} = (1-f_{\rm d}),
\end{equation}

\begin{equation}\label{stoppingtime}
    t_{\rm s} = \frac{\rho_{\rm d}\rho_{\rm g}}{K(\rho_{\rm d}+\rho_{\rm g})},
\end{equation}
We note that our stopping time definition is the same as in \cite{lin17}; however, it is different to particle stopping time $\tau_{\rm s}$ used in other studies e.g. \citealt{youdin05}) 
$\tau_{\rm s} = t_{\rm s}\rho/\rho_{\rm g}$ and the 'relative' stopping time, $t_{\rm s}$ is related to the Stokes number through,

\begin{equation} \label{stokesnum}
    {\rm St} = \Omega_{\rm K}t_{\rm s}\frac{\rho}{\rho_{\rm g}}.
\end{equation}
It is worth noting that within FARGO3D, a dust fluid is characterised by a constant Stokes number.

Using the identities (\ref{gasvel})-(\ref{stoppingtime}), we can rewrite the two fluid equations in terms of the barycentric frame,
\begin{align}
\label{mixcont}
    \frac{\partial\rho}{\partial t} + \nabla\cdot(\rho\textbf{u}) =& 0,\\
    \frac{\partial\textbf{u}}{\partial t} + (\textbf{u} \cdot\nabla)\textbf{u} +\frac{\nabla P_{\rm g}}{\rho} + \frac{1}{\rho} \nabla\cdot [f_{\rm d}\rho(1-f_{\rm d})\Delta\textbf{u}\Delta\textbf{u}] =& 0,\\
    \frac{\partial{f_{\rm d}}}{\partial t} + (\textbf{u}\cdot\nabla)f_{\rm d} + \frac{1}{\rho} \nabla\cdot [f_{\rm d}\rho(1-f_{\rm d})\Delta\textbf{u}] =& 0,\\
\label{mixrel}
    \frac{\partial\Delta\textbf{u}}{\partial t} + (\textbf{u}\cdot\nabla)\Delta\textbf{u} + (\Delta\textbf{u}\cdot\nabla)\textbf{u} +&\nonumber\\ \frac{\Delta\textbf{u}}{t_{\rm s}} - \frac{\nabla P_{\rm g}}{(1-f_{\rm d})\rho} - \frac{1}{2}\Delta[(2f_{\rm d} - 1)\Delta\textbf{u}^2] =& 0
\end{align}

\subsection{Locally Isothermal Terminal Velocity Approximation (LITVA)}

In the strong drag regime where ${\rm St}\ll1$, the stopping time $t_{\rm s}$ is small, reaching a point where the relative velocity is heavily damped. This is through $\Delta\textbf{u}/t_{\rm s}$ from equation (\ref{mixrel}) as the dust velocity adjusts to the gas phase quickly. We can then use the "terminal velocity approximation" whereby equation (\ref{mixrel}) simply reduces down to

\begin{equation} \label{tvamixrel}
    \Delta\textbf{u} = \frac{\nabla P_{\rm g}}{\rho_{\rm g}} t_{\rm s}.
\end{equation}

Additionally, the terms $\Delta\textbf{u}\Delta\textbf{u}$ can also be neglected in this approximation which simplifies our single fluid equations down to first order approximation
\begin{align}
\label{mixcont2}
    \frac{\partial\rho}{\partial t} + \nabla \cdot (\rho\textbf{u}) =& 0,\\
\label{mixvel2}
    \frac{\partial \textbf{u}}{\partial t} + \textbf{u}\cdot\nabla\textbf{u} =& \nabla\Phi - \frac{\nabla P_{\rm g}}{\rho},\\
\label{dustfracevo}
    \frac{\partial {f_{\rm d}}}{\partial t} + \textbf{u}\cdot\nabla{f_{\rm d}} =& -\frac{1}{\rho} \nabla\cdot(f_{\rm d} t_{\rm s} \nabla P_{\rm g}).
\end{align}
Since we consider a locally isothermal disc, using $P_{\rm g}=c_{\rm s}({\bf r})^2 \rho_{\rm g}$, we can therefore reformulate equation (\ref{dustfracevo}) through rearranging the equation of state with equation (\ref{gasfrac}) as a dust fraction (\citealt{lin17}),

\begin{equation}\label{dustfracpressure}
    f_{\rm d} = 1 - \frac{P_{\rm g}}{{c}_{\rm s}^2(r,z)\rho},
\end{equation}

\noindent leading to,

\begin{equation} \label{pressevo}
    \frac{\partial P_{\rm g}}{\partial t} + \nabla \cdot (P_{\rm g}\textbf{u}) = P_{\rm g}\textbf{u}\cdot\nabla \ln {c}_{\rm s}^2 + C,
\end{equation}
where $C$ is the cooling term of the dust:
\begin{equation}\label{coolingterm}
    C = {c}_{\rm s}^2\nabla\cdot \Bigg[t_{\rm s}\Bigg(1-\frac{P_{\rm g}}{{c}_{\rm s}^2\rho}\Bigg)\nabla P_{\rm g}\Bigg].
\end{equation}

In the implementation of the single fluid model in the strong drag regime, we will only need to solve equations (\ref{mixcont2}), (\ref{mixvel2}) and (\ref{pressevo}). The overall advantage of this simplification is that fewer equations need to be solved hence less computational cost is needed. Additionally, the back reaction of the dust on the gas is intrinsically included through the relative velocity evolution that has been simplified under the terminal velocity approximation and there are no severe time step constraints that come with explicitly evaluating the drag between the dust and gas for small stopping times.

\subsection{2D Thin Discs}

The equations presented above are in 3D, however in our simulations we assume a geometrically thin disc. Therefore we need 2D cylindrical equations which are derived from vertical integration of the quantities above to $(r,\phi)$. The density $\rho$ is replaced with surface density for the mixture, denoted by $\Sigma$ and the pressure $P_{\rm g}$ with vertically integrated pressure, $p_{\rm g}$ . The continuity and momentum equations are then changed to,
\begin{align}
    \frac{\partial\Sigma}{\partial t} + \nabla \cdot (\Sigma\textbf{u}) =& 0,\\
    \frac{\partial \textbf{u}}{\partial t} + \textbf{u}\cdot\nabla\textbf{u} =& \nabla\Phi - \frac{\nabla p_{\rm g}}{\Sigma},
\end{align}
\noindent with the potential being softened over a length scale $0.6H$ with $H$ as the scale height and evaluated at the midplane, $z = 0$. For the pressure evolution we are solving,
\begin{equation} \label{}
    \frac{\partial p_{\rm g}}{\partial t} + \nabla \cdot (p_{\rm g}\textbf{u}) = p_{\rm g}\textbf{u}\cdot\nabla \ln {c}_{\rm s}^2 + C,
\end{equation}
with
\begin{equation}\label{}
    C = {c}_{\rm s}^2\nabla\cdot \Bigg[t_{\rm s}\Bigg(1-\frac{p_{\rm g}}{{c}_{\rm s}^2\Sigma}\Bigg)\nabla p_{\rm g}\Bigg].
\end{equation}
In other words, the density and pressure are replaced in the cooling term with the vertically integrated quantities. We can not consider dust settling, $v_{\rm z}$, in the vertically integrated equations and as the cooling term is an effective heating/cooling effect via the transportation of dust through the flow of the gas, there is no cooling effect in the vertical direction.

\section{Method} \label{method}

The simulations presented use both the modified version of FARGO with LITVA implemented and FARGO Multifluid where multiple dust species can be evolved separately with the gas. We will be  testing our modification of a single fluid mixture and a Two Fluid model with gas and one species of dust. 

Within the code, we adopt dimensionless units whereby distance is measured in terms of the orbital radius of the planet (which does not migrate), $r_{\rm p}$, one orbital period is $2\pi$ and mass is measured in terms of the central mass.

The radial domain of the simulations is from 0.4 $r_{\rm p}$ to 2.5 $r_{\rm p}$ (\citealt{devalborro06}) with a constant disc aspect ratio $h_0=H/r = 0.05$ with $H$ as the scale height. We note that there is no flaring of the disc. The lowest resolution in the tests are 256 and 512 uniformly spaced cells in the radial and azimuthal direction respectively. This is used for the axisymmetric case as the disc morphology is not significantly affected whilst further results increases the resolution for comparison and clarification of substructures in the presence of planets especially in the horseshoe regions. In \cite{lovascio19}, viscosity was included in the LITVA model and we use a uniform kinematic viscosity value of $\nu = 10^{-5}$ with units $r^2_{\rm p} \Omega_k(r_{\rm p})$ which corresponds to an $\alpha$ viscosity value of $4\times 10^{-3}$ from \cite{shakura73}, where $\nu = \alpha c_{\rm s} H$ . For the Two Fluid model the gas and dust surface densities are set to,
\begin{align}
    \Sigma_{\rm g} =& \Sigma_0\Bigg(\frac{r}{r_{\rm p}}\Bigg)^{-0.5}(1-f_{\rm d}),\\
    \Sigma_{\rm d} =& \Sigma_0\Bigg(\frac{r}{r_{\rm p}}\Bigg)^{-0.5}f_{\rm d},
\end{align}
with $\Sigma_0$ as the initial total surface density of the gas and dust at the location $r_{\rm p}$. The dust fraction $f_{\rm d}$ is constant, initially. Since we do not consider self-gravity the value of $\Sigma_0 = 6.3662\times10^{-4}$ from FARGO3D default models is used in both models as it provides no additional physical effects. 

To test the regime of strong drag we vary the constant Stokes number used in FARGO3D through the stopping time in equation (\ref{stokesnum}) for the single fluid model. We use the barycentric values for the gas and the dust, setting the single fluid mixtures surface density as,

\begin{equation}
    \Sigma = \Sigma_0\Bigg(\frac{r}{r_{\rm p}}\Bigg)^{-0.5},
\end{equation}

\noindent while the initial pressure profile of the disc is reduced by the dust fraction,

\begin{equation}
    p_{\rm g} = c_{\rm s}^2(r)\Sigma_{\rm g} = c_{\rm s}^2(1-f_{\rm d})\Sigma,
\end{equation}

\noindent with the sound speed as $c_{\rm s}(r) = h_0\Omega_{\rm K}r$ and the azimuthal and radial velocities as the barycentric velocities of the mixture. The dust and gas in the single fluid model start off with a barycentric radial velocity of zero, however they have their own initial radial drift due to the pressure gradient. This is due to the the combination of the identities and the terminal velocity approximation, (\ref{gasvel}), (\ref{dustvel}) and (\ref{tvamixrel}) which we present in 2D,

\begin{equation} \label{initgasvel}
    \textbf{u}_{\rm g,0} =  - \frac{\Sigma_{\rm d}}{\Sigma} \frac{\nabla p_{\rm g}}{\Sigma_{\rm g}} t_{\rm s}
\end{equation}

\begin{equation}\label{initdustvel}
    \textbf{u}_{\rm d,0} =   \frac{\nabla p_{\rm g}}{\Sigma} t_{\rm s}
\end{equation}

The modifications we added to FARGO3D build upon \cite{lovascio19}, where the TVA approximation is implemented with the pressure evolution equation replacing the energy evolution in the code and the cooling term is evolved using an RKL2 method (\citealt{meyer14}) for stability. We modify the code further to include cylindrical coordinates to simulate a global disc and a locally isothermal equation of state. The Stockholm Boundary (\citealt{devalborro06}) is implemented for the pressure evolution at the radial boundaries of the disc. These are wave-killing zones to damp disturbances near the mesh boundaries. The boundary conditions for azimuthal velocity are an extrapolation of the Keplarian profile. For density and pressure it is extrapolated using its initial power law profile and an anti-symmetric boundary condition is applied to the radial velocity. 

\section{Results} \label{zero}

In the following subsections we present the results of our locally isothermal terminal velocity approximation modification and compare them to the full two-fluid model. The section \ref{test} demonstrates the agreement between the two models for an axisymmetric disc, in section \ref{low} we embed the disc with an Earth mass planet and test the limiting Stokes number in \ref{stokes}. In \ref{large} we test the limits of the planet mass that can be embedded in the disc for this modification. Finally we compare the computational costs of both models in subsection \ref{cost}.

\subsection{Axisymmetric Disc}\label{test}

We run initial tests of the gas and dust as single fluid in an axisymmetric disc with no planets embedded. From \cite{nakagawa86}, the expectation is that the dust drifts radially inward due to the gas travelling at sub-Keplerian speeds due to the pressure. This causes a drag on the dust causing the dust to lose angular momentum. For the test case we consider the Stokes number of the dust St$ = 0.01$. The dust to total density considered $f_{\rm d} \in [0.01,0.1]$. We choose a dust to total ratio of $1\%$ for typical composition of a protoplanetary disc and 10\% to test any effects of a higher dust to gas ratio. We use a resolution of 256 cells in the radial direction respectively to compare with global disc simulations of Two Fluid approach and the simulation time is 150 orbits.

\begin{figure}
\includegraphics[scale=0.38]{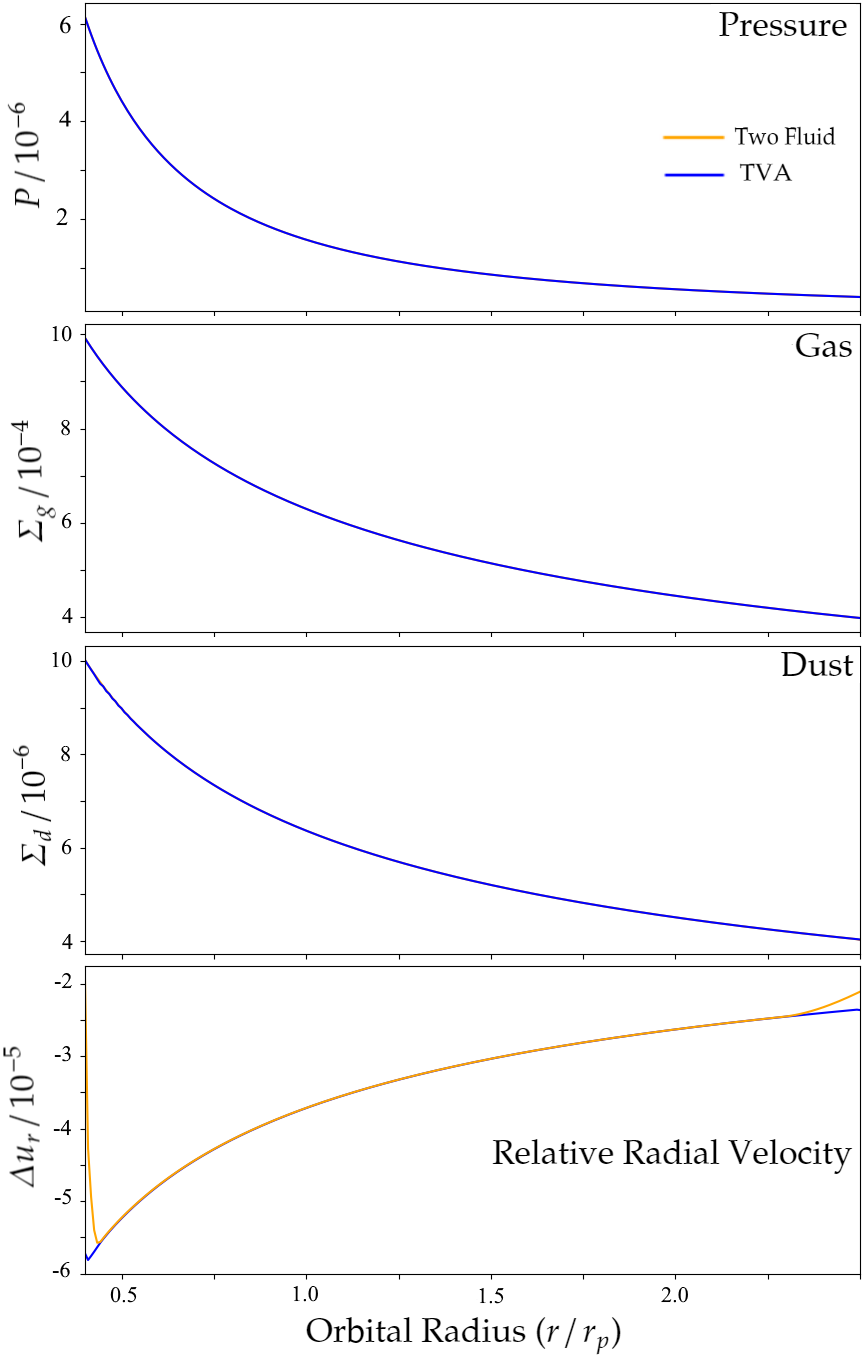}
\caption{Pressure, dust surface density, gas surface density and relative radial velocity between the gas and dust for an axisymmetric disc with St$ = 0.01$ and $f_{\rm d} = 0.01$ with no planets embedded. The blue and orange curves correspond to the TVA and Two Fluid models respectively.}
\label{fig:zerocomps}
\end{figure}

In Fig. \ref{fig:zerocomps} we compare the pressure and densities from both models. For the TVA model we calculate the values of the dust and gas density from the pressure and total density of the mixture at each point using equation \ref{dustfracpressure} and the locally isothermal equation of state. We can see that for the first three plots of Fig. \ref{fig:zerocomps} the two models overlap very well and agree precisely. The fourth plot compares the relative radial velocity between the gas and dust, for both models. For the Two Fluid model we plot the difference between their radial velocities and for the TVA model we calculate the relative radial velocity using equation \ref{tvamixrel}. Again we see a close agreement between the two models. The Two Fluid model adjusts to this equilibrium solution very quickly from zero initial velocity for the dust and gas. The differences appear from the boundaries where the FARGO Two Fluid implementation cuts off the radial drift of the dust and gas compared to the TVA model where the radial velocities continue smoothly to the edges of the disc. This is due to the Stockholm boundary implementation (\citealt{devalborro06}) in FARGO3D as the Two Fluid simulations do not start in a state of equilibrium. To conclude, for a single fluid mixture under the terminal velocity approximation with dust of Stokes number of $10^{-2}$ and dust fraction of 1\% the two models agree well for an axisymmetric disc with no planets embedded for the pressure, densities and velocities evolution.

\subsubsection{Increasing Dust Fraction}

\begin{figure}
\includegraphics[scale=0.38]{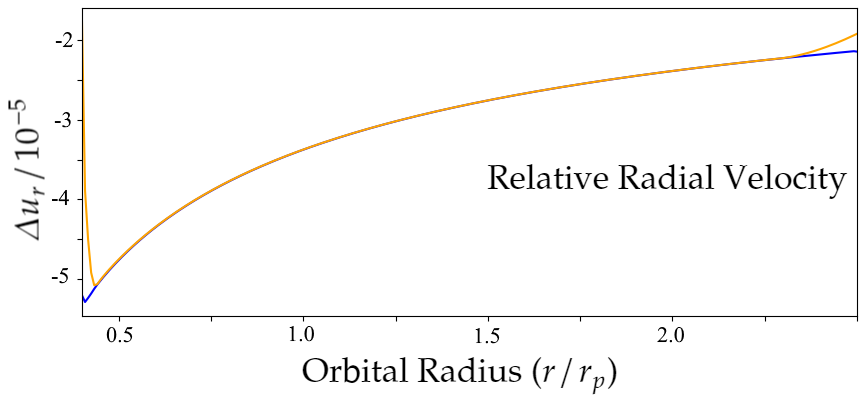}
\caption{Relative radial velocity between the gas and dust for a dust ratio $f_{\rm d} = 0.1$ and St$=0.01$. The blue and orange curves correspond to the TVA and Two Fluid models respectively.}
\label{fig:zero01rel}
\end{figure}

Before placing a planet in the disc and testing higher sized dust grains through the Stokes number, we extend the test case of an axisymmetric disc for a higher dust to total ratio of 1:10. Fig. \ref{fig:zero01rel} presents the azimuthally averaged relative radial velocity between the dust and gas for our higher dust fraction. The pressure and densities were identical as with Fig. \ref{fig:zerocomps} that the TVA model agrees precisely at 150 orbits with Two Fluid and we have only included the relative radial velocity to show that the cooling term, see equation. \ref{coolingterm}, which represents the drift of dust to be evolving correctly for a higher dust content in the disc. This will be the limit of the dust to gas ratio tested in this case but we will explore higher ratios when including planets as we are interested in not only the evolution of higher dust densities around a planet but also the view that as dust collects in pressure bumps which can be caused by planet disc interactions, the dust to gas ratio naturally increases.

In both test cases for an axisymmetric disc with Stokes number $10^{-2}$ we find that the evolution of the dust and gas mixture agree very well with the dust and gas evolved separately for a 1:100 and 1:10 dust to total ratio. This follows the strong drag regime and we have shown that the implementation of cylindrical coordinates and locally isothermal equation of state for the TVA model for an axisymmetric disc is accurate. 

\subsubsection{Limiting Stokes Number for Axisymmetric Disc}

In this subsection we start to test the limits of the model in terms of increasing Stokes number. Firstly as we know the terminal velocity approximation is accurate when considering the strong drag regime between the dust and gas, to test the limits we consider a larger Stokes number than the previous St$=0.01$. This increase in Stokes number links to a weaker drag and in turn, the dust becomes less coupled to the gas. We consider the case for an axisymmetric disc with no planet embedded. The Stokes numbers we consider are St $\in [0.1,0.2,0.5]$ and present our results along with our first test cases of agreement.

\begin{figure}
\includegraphics[scale=0.36]{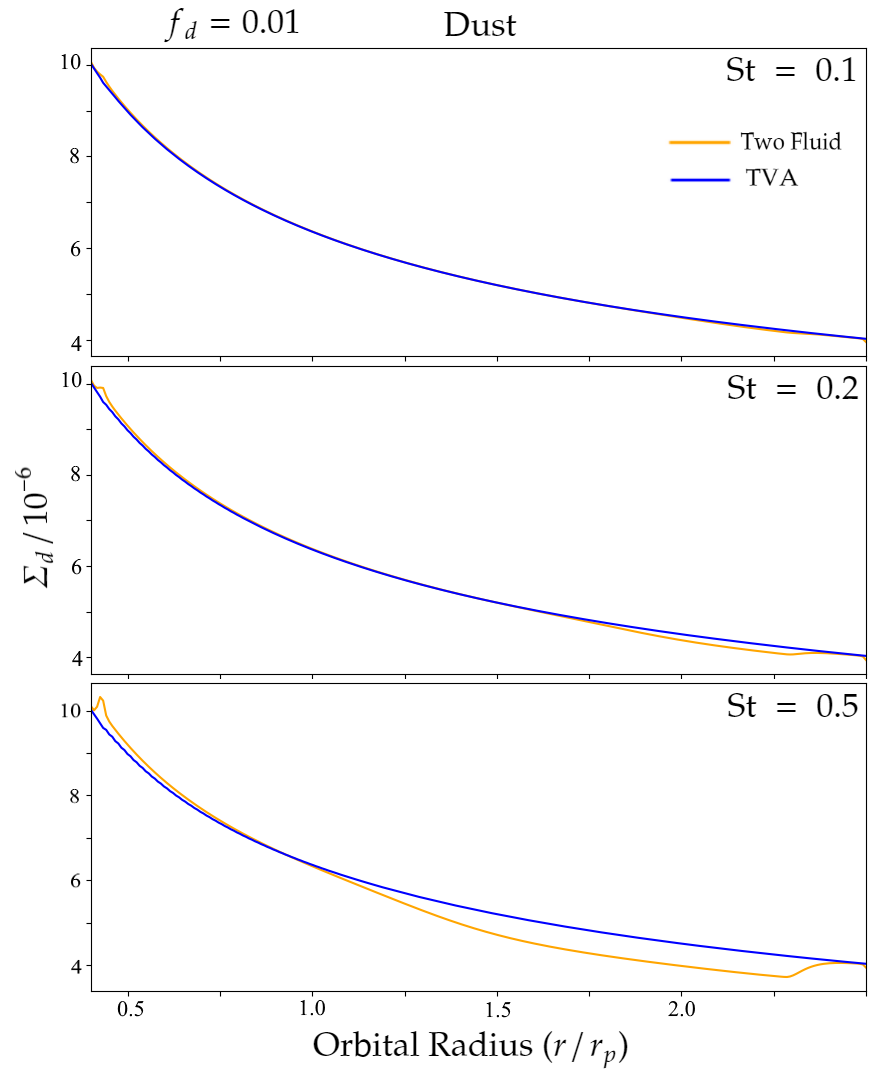}
\caption{Dust surface density for increasing Stokes numbers, St$ = 0.1, 0.2, 0.5$ and $f_{\rm d} = 0.01$ with no planets embedded.}
\label{fig:dusthighst}
\end{figure}

For the axisymmetric disc with no planet embedded, Fig. \ref{fig:dusthighst} shows the azimuthally averaged dust surface density for increasing Stokes number for the TVA model in comparison to Two Fluid. Immediately we see that as the Stokes number is increased, there is a discrepancy between the two models. For the Two Fluid model, the dust buildup interior is due to the dust drifting inwards but at the boundaries, the radial velocities are cut off as shown in the differences for Fig. \ref{fig:zerocomps}. This would also therefore cause the depression towards the outer boundary for the surface density as the dust drifted in. In the LITVA model the boundaries allow a stable evolution for the dust surface density over a long period.

\begin{figure}
\includegraphics[scale=0.38]{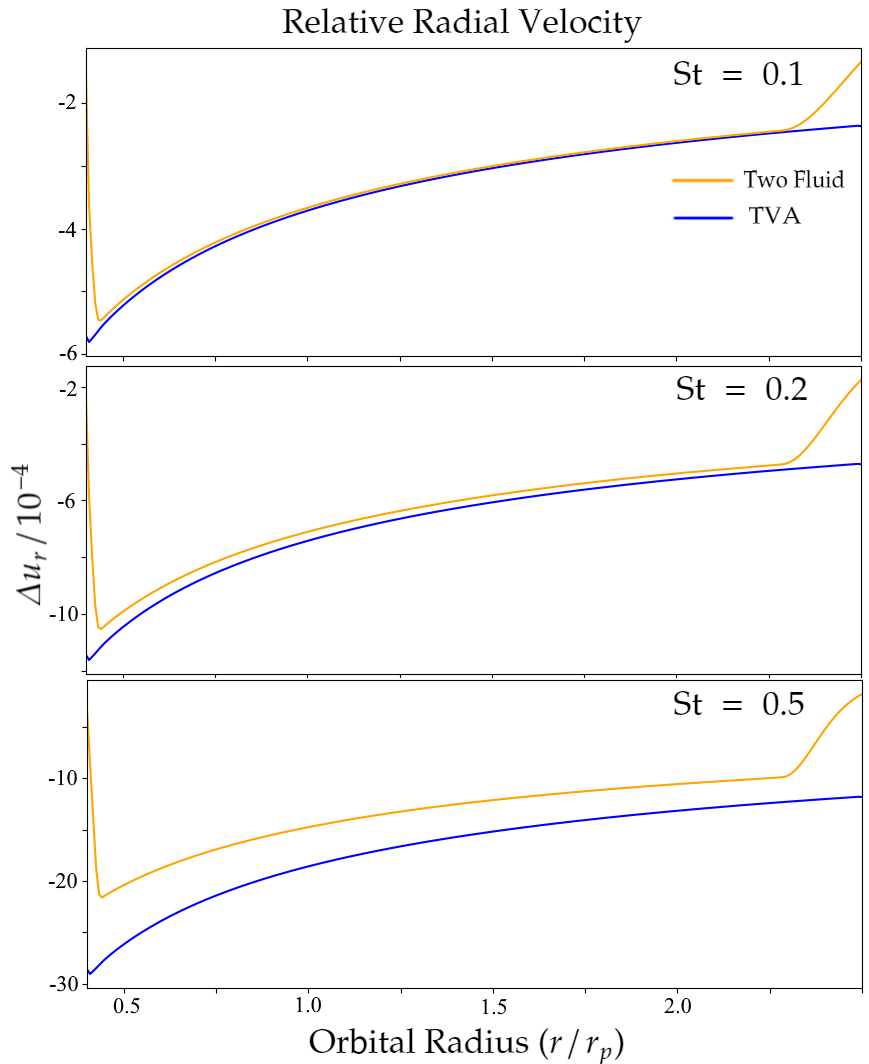}
\caption{Relative radial velocity for increasing Stokes numbers, St$ = 0.1, 0.2, 0.5$ and $f_{\rm d} = 0.01$ with no planets embedded.}
\label{fig:relhighst}
\end{figure}

In Fig. \ref{fig:relhighst} we show the calculated azimuthally averaged radial velocity between the two phases and see that the TVA model does not match with the Two Fluid model as expected as we are departing from the strong drag regime with an increasing Stokes number. In both evolutions of the dust density and velocities, we argue that for the axisymmetric disc that the limiting constant Stokes number for the regime of strong drag under our setup and model would be around 0.1. Higher values would seem unreasonable to model with based on the results for the well coupled regime as the relative radial velocities between the dust and gas starts to diverge.

\subsection{Earth Mass Planet}\label{low}

Having shown that the TVA single fluid model agrees well with the Two Fluid model, we consider how the model holds up with the presence of a planet in the disc. In this section we test an Earth mass planet, $3\times10^{-6}M_{P}$ embedded in the disc where $M_P$ is the mass of the primary. Our expectations from this setup are that the Earth mass planet would not produce shocks that lead to a break down of the TVA model as shown in \cite{lovascio19} where a first order error is produced around shocks compared to a two fluid model. We expect the two models to be in good agreement in terms of the evolution of the disc. For this test we keep the same Stokes number St$ = 0.01$ and dust fractions $f_{\rm d} \in [0.01,0.1]$. The resolution is (384,768), corresponding 6 cells per scale height at the planet's location. A simulation time of 500 orbits was used to compare the longer term evolution of the disc with a planet present and the substructures seen. Later on in Section \ref{large}, we test higher planet masses up to Neptune sized.

\begin{figure}
\includegraphics[scale=0.37]{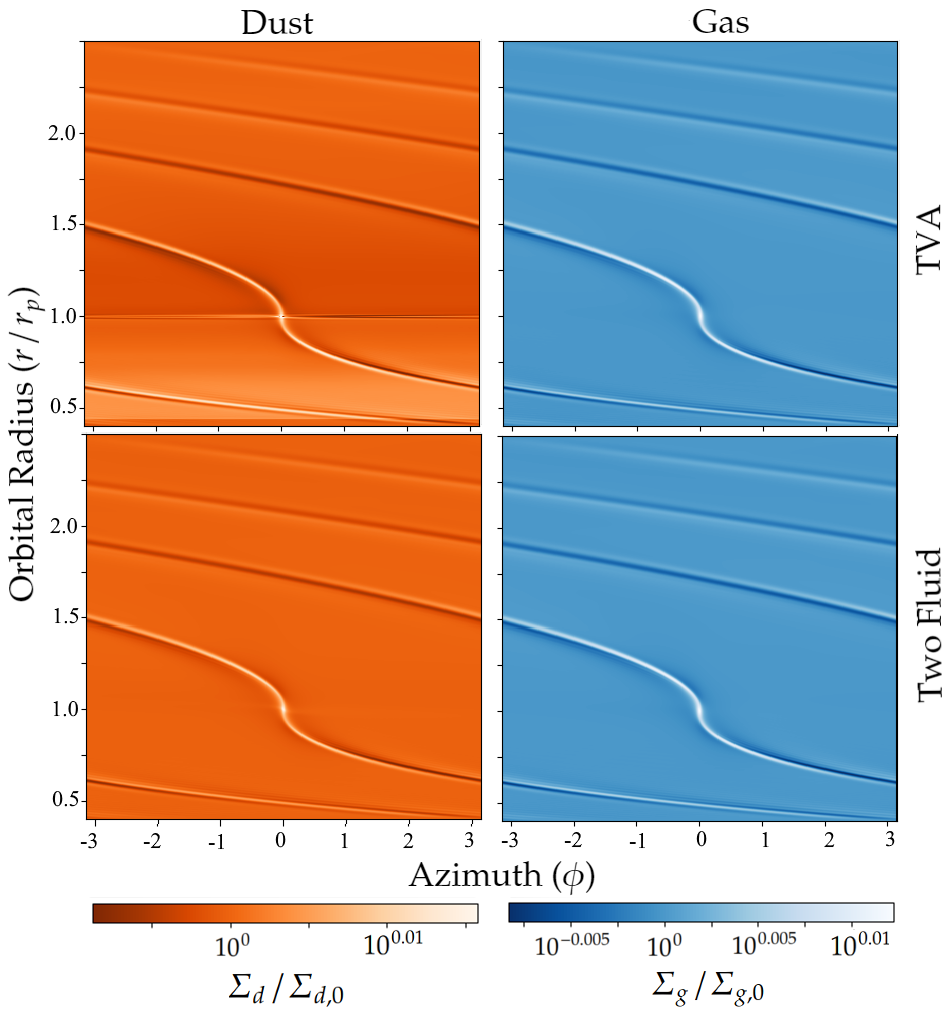}
\caption{Dust (left column) and gas (right column) surface density plots for the TVA (top row) and Two Fluid (bottom row) models. The disc is embedded with an Earth mass planet at 1 $r_{\rm p}$. The time period $T_p = 500$ orbits, Stokes number St $=0.01$ and dust fraction $f_{\rm d}=0.01$.} 
\label{fig:earthcomps001}
\end{figure}

Fig. \ref{fig:earthcomps001} shows the dust and gas surface density plots for both the TVA and Two Fluid model with dust fraction of $0.01$. We can see that in both models, for both the dust and gas plots, the Earth sized planet has induced a spiral density wave that propagates through the disc. The launching of density waves through the disc by a planet has been well studied since the early works of Goldreich \& Tremaine (\citeyear{goldreich79}, \citeyear{goldreich80}). We can see that this one armed spiral density wave generated by our Earth sized planet in both models agree with the expectation of the wake generated by a low mass planet (\citealt{ogilvie02}) and the disc morphology has not changed significantly due to the planet. There are however differences between the two models in which the TVA model plots have shown a substructure in the horseshoe region of the planet. Additionally there is an $\approx 1\%$ dust buildup towards the interior of the planet.

\begin{figure}
\includegraphics[scale=0.37]{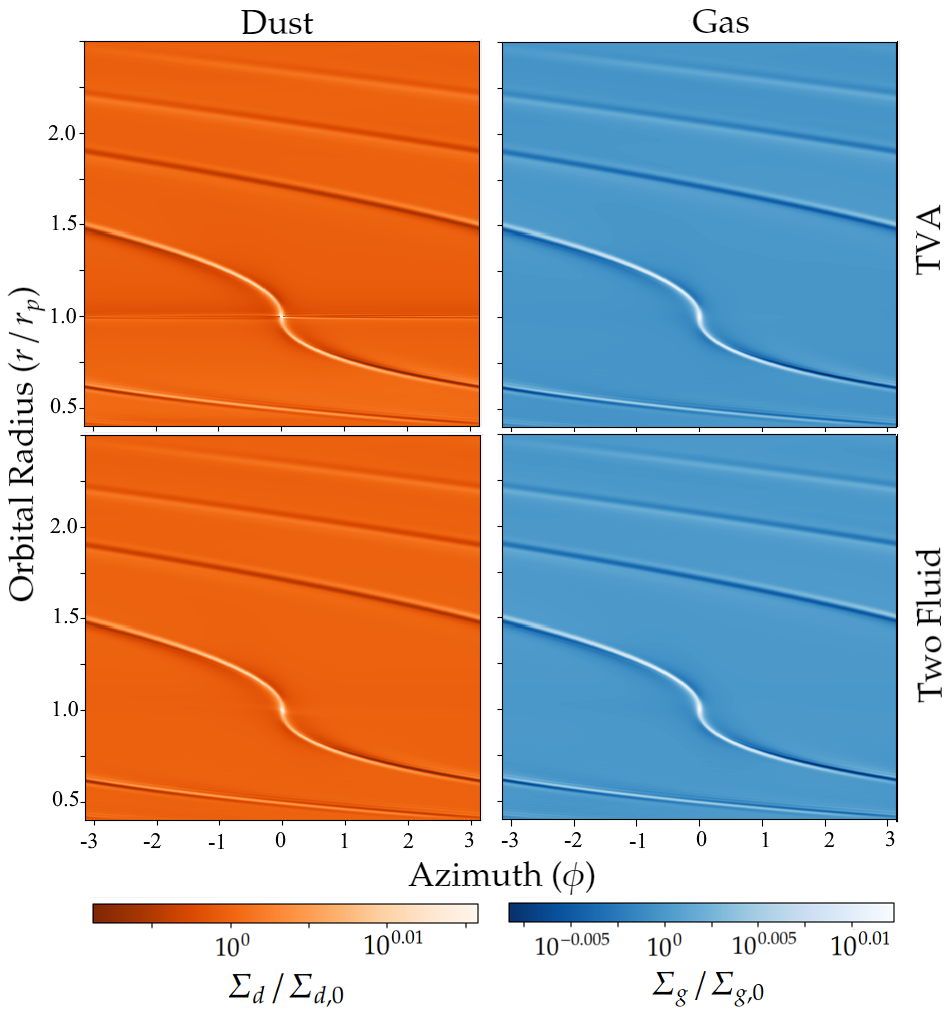}
\caption{Dust (left column) and gas (right column) surface density plots for the TVA (top row) and Two Fluid (bottom row) models.. The disc is embedded with an Earth mass planet at 1 $r_{\rm p}$. The time period $T_p = 500$ orbits, Stokes number St$=0.01$ and dust fraction $f_{\rm d}=0.05$.} 
\label{fig:earthcomps005}
\end{figure}

In Fig. \ref{fig:earthcomps005} we increase the dust fraction to $5\%$ and again we see that there is a ring like substructure in the horseshoe region. Noticeably the inner dust buildup is not present for the higher dust fraction. We believe that the growth of dust towards in the inner boundary may be due to the first term on the right hand side of equation (\ref{pressevo}). In a previous study by \cite{mcnally19a}, whereby a gas only disc was implemented in FARGO3D, an equivalent temperature effect was seen in the inner disc, which could be attributed to the PdV work source term. By solving for the specific entropy rather than the internal energy, this feature could be avoided \citep{mcnally19a}.   
We believe this is the same case for this model where the pressure is evolved in the place of the energy equation due to our source term for the pressure evolution stemming from the locally isothermal nature of our disc. In order to fix this issue, we would have to change the evolution equation in future so that the source term disappears. This would be through a transformation of the pressure evolution using $P_{\rm g}=c_{\rm s}^2 \rho_{\rm g}$ where the gas fraction is evolved instead. However, in this paper, we focus on the overall structure of the disc and most importantly the interactions around the planet rather than the disc's interaction at the boundaries as they do not affect the disc morphology significantly.

In our results for the disc embedded with an Earth mass planet, the ring like structure around the horseshoe region is stable over a long evolution. Although this feature is not seen in the Two Fluid model results, previous studies including the LITVA model (\citealt{chen18}) and low viscosity or inviscid disc (e.g. \citealt{dong17}; \citealt{hsieh20}) have shown ringed structures around low mass planets. In \cite{chen18}, they implement the same dynamical model of a single fluid mixture into PLUTO with a $M = 10M_{\earth}$ planet, resulting in a multi-ringed structure associated with the gas gap edges and horseshoe region of the planet (see their Appendix A). The difference in planet mass embedded could explain the clearer ring structure in their results compared to ours as a shallower and wider gap is carved out over a time period of 700 orbits. In \cite{dong17}, they present a two fluid model using LA-COMPASS code showing again the formation of a multi-ringed structure in the horseshoe region of a $M = 10M_{\earth}$ planet (see their Fig. 1). When it comes to a more similar setup with ours of lower mass planets, \cite{hsieh20} incorporates a $M = 2M_{\earth}$ in a Two Fluid model whereby in a low metallicity and Stokes number simulation, substructures of rings in the horseshoe region are produced similar to the ones we see in our results. They provide plots for a migrating planet and azimuthally averaged surface density in the dust for both migrating and non-migrating in their Figures 1 and 4. Therefore the presence of these ring like structures in the horseshoe region of a low mass planet embedded is not surprising. Their absence in the two fluid model means that either they are an artifact of LITVA, or that the resolution constraints on the two fluid model are too severe for this feature to appear.

\subsubsection{Corotation dust ring} \label{differences}

\begin{figure}
\centering
\includegraphics[scale=0.39]{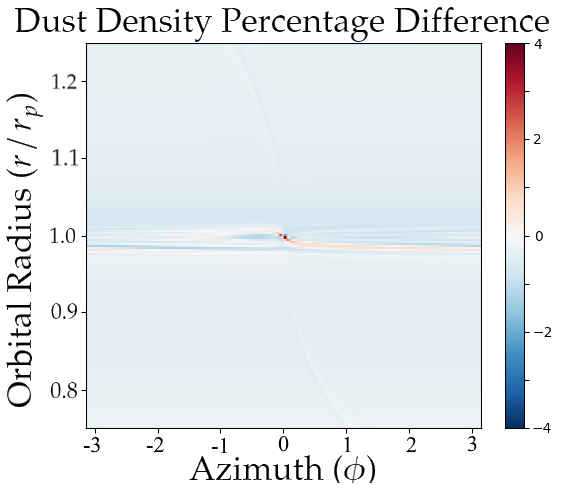}
\caption{ Dust density percentage difference between the TVA and Two Fluid model. The disc is embedded with an Earth mass planet at 1 $r_{\rm p}$. The time period $T_p = 2000$ orbits, Stokes number St$ = 0.01$ and dust fraction $f_{\rm d}=0.01$.} 
\label{fig:dustperc}
\end{figure}

We first test the robustness of the feature seen at corotation in the TVA model by doubling the resolution to (768,1536), corresponding to 12 cells per scale height at the planet's location, and run the simulation for 2000 orbits. The model parameters are otherwise the same as before with a Stokes number, ${\rm St} = 0.01$ and dust fraction, $f_{\rm_{d}} = 0.01$. The model reaches the same configuration in terms of the dust density as seen previously, with a dust feature present in the horseshoe region and in Fig. \ref{fig:dustperc} we present the dust density percentage difference between the two models after 2000 orbits.

We see that the dust density, relative to the two fluid model, differs between the two models with a maximum reduction in dust density of $-2.27\%$ and a maximum enhancement of dust density of $+3.91\%$ nearby the planet and an average of around $1-2\%$ in the horseshoe region itself with the dust ring. Additionally we calculated the percentage difference at 500, 1000 and 1500 orbits for the dust density which showed a maximum reduction of $-2.00,-2.14,-2.22\%$ and enhancement of $+4.19,+4.05,+3.96\%$ respectively. The evolution of the percentage differences between the model at different times indicates that the configuration is stable in terms of the density evolution of the dust over long periods after the feature appears near early evolution of the disc. We note that the gas density percentage difference in both 1000 and 2000 orbits in reduction and enhancement were around $\pm0.02\%$ which affirms that the gas density distribution is largely identical between the two models. Based on the dust density itself we can conclude that the overall morphology for long periods and higher resolution does not differ much between the two models. 

 Both the TVA and the two-fluid model therefore appear to be converged with resolution. In other words, the dust ring does not begin to show up in the two fluid model at higher resolution, nor does it start to disappear in the TVA model. The TVA is known to introduce unphysical behaviour in some cases, such as spurious modes in streaming instability calculations \citep{lin17}. It is therefore possible that this dust ring is an artifact of the TVA formulation. Another option is that these rings are a dusty equivalent of thermal lobes \citep{2017MNRAS.472.4204M}, or the 'cold finger' \citep{lega2014}, which we will discuss in more detail in section \ref{sec:lobes}.

\begin{figure}
\centering
\includegraphics[scale=0.21]{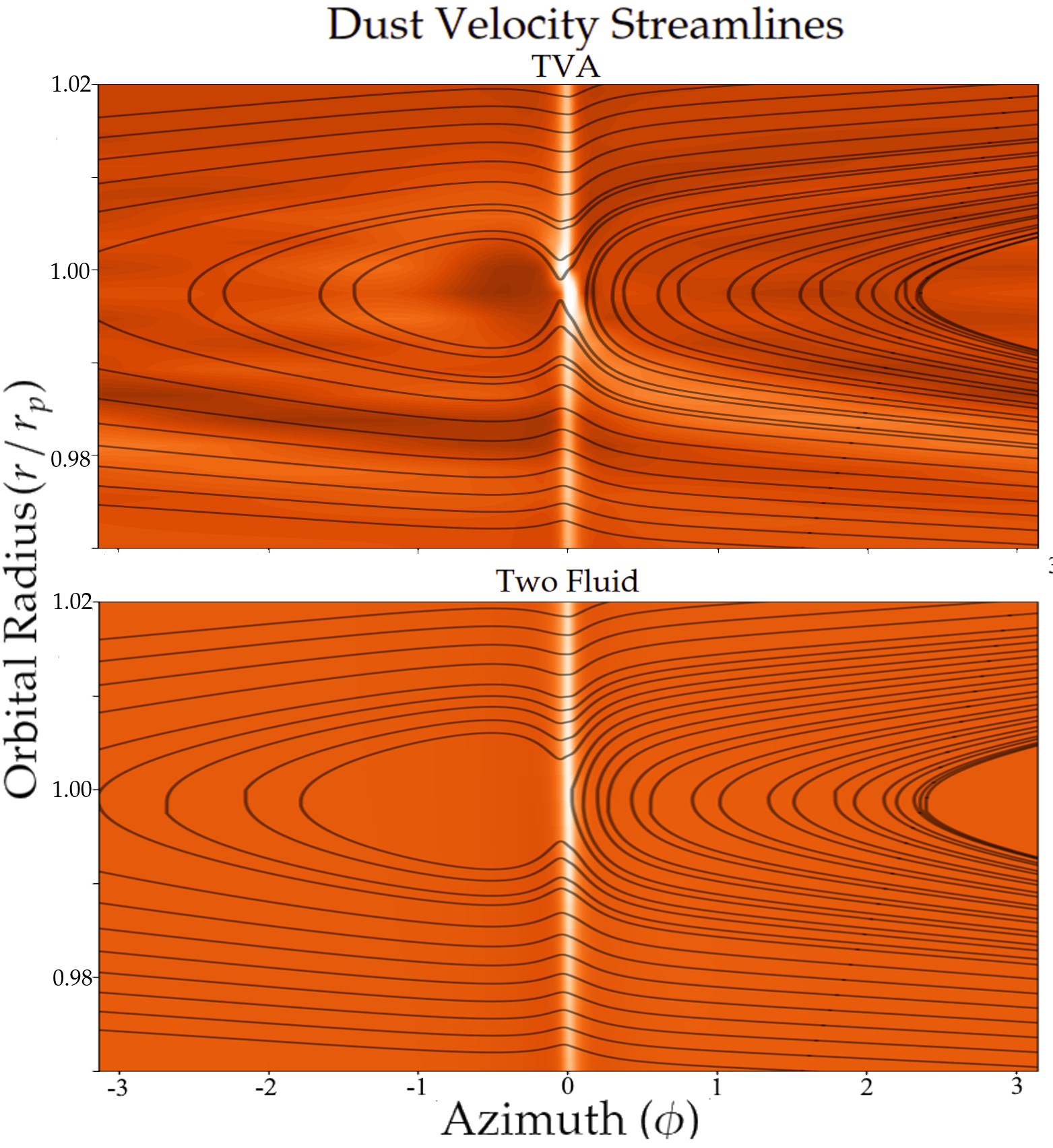}
\caption{ Dust velocity streamlines for TVA and Two Fluid model. The time period $T_p = 2000$ orbits, Stokes number St$ = 0.01$ and dust fraction $f_{\rm d}=0.01$.} 
\label{fig:duststream}
\end{figure}

 Next we compare the dust velocity streamlines near the planet in Fig \ref{fig:duststream}. We see that the flow structure of the dust in the artifact region of the TVA model follows expectations of an asymmetry between the two horseshoe legs due to a migrating planet \citep{masset2003}, but in our case it is due to the radial drift of dust as the planet is in a fixed orbit. Comparing the two models, we see that the streamlines match up well with minor differences. The dust ring in the TVA model does not change the shape of the streams over a long period and keeps the expected shape in the horseshoe region of the planet.

\begin{figure*}
\includegraphics[scale=0.40]{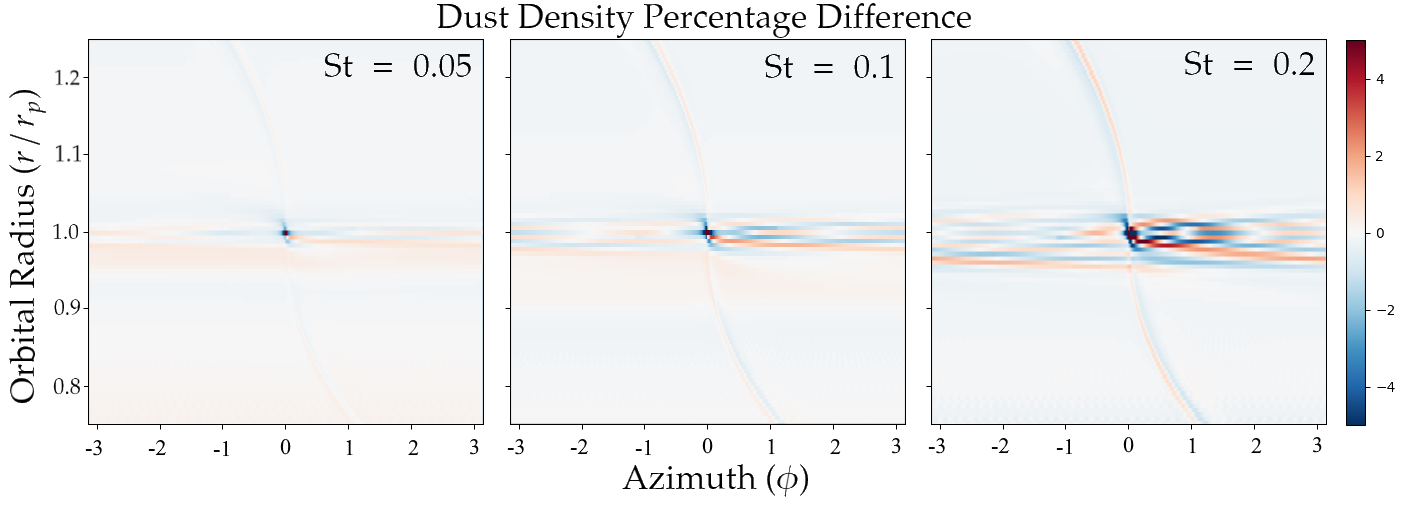}
\caption{Percentage difference of TVA model compared to the Two Fluid model for dust surface density. The colour bar is limited to $\pm5\%$ difference to show a general pattern of how the differences are spread out as a whole picture rather than specifically how much percentage difference over or under $\pm5\%$ for each cell. The time period $T_p = 500$ orbits, Stokes number St$ \in [0.01,0.05,0.1,0.2]$ and dust fraction $f_{\rm d}=0.01$.} 
\label{fig:densdiff}
\end{figure*}

\begin{figure}
\includegraphics[scale=0.37]{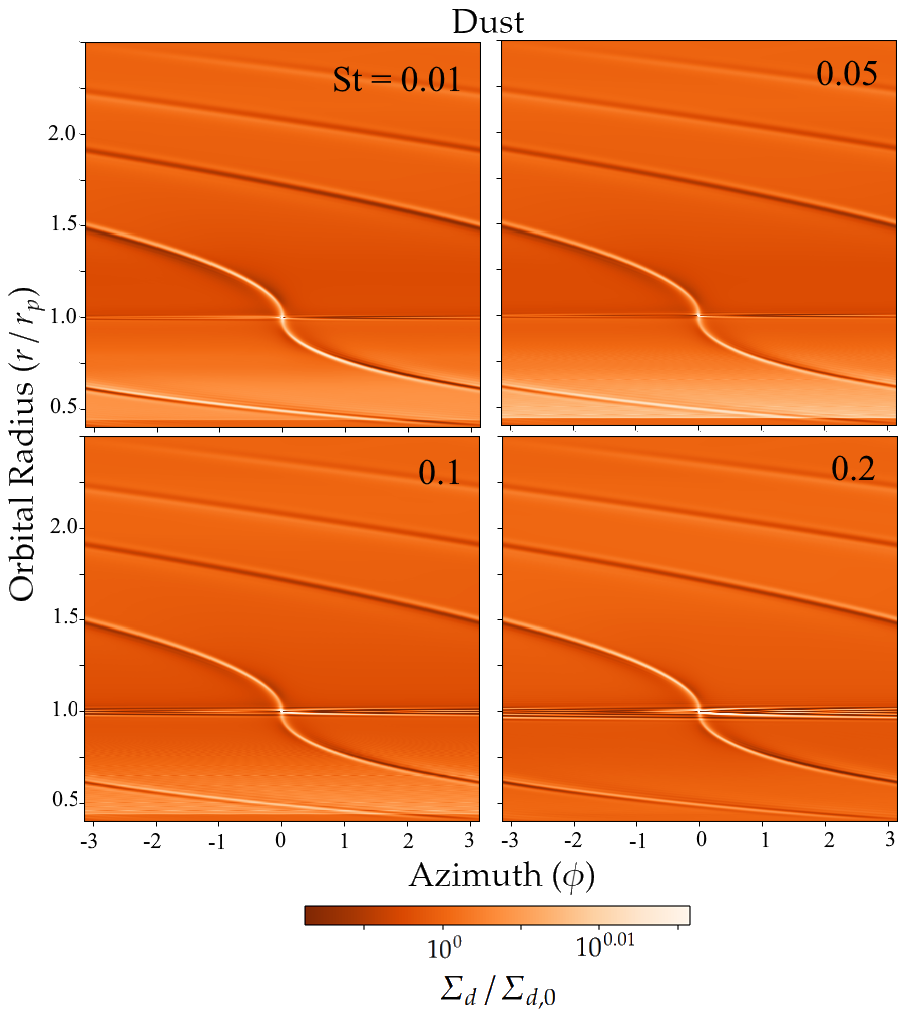}
\caption{Dust surface density plots for TVA model. The disc is embedded with an Earth mass planet at 1 $r_{\rm p}$. The time period $T_p = 500$ orbits, Stokes number St$ \in [0.01,0.05,0.1,0.2]$ and dust fraction $f_{\rm d}=0.01$.} 
\label{fig:earthhigherst}
\end{figure}

\subsubsection{Limiting Stokes Number for Earth Mass Planet}\label{stokes}

In the case of an Earth mass planet embedded, we consider Stokes numbers, St $\in [0.05,0.1,0.2]$ to compare with our test case of $0.01$ in Section \ref{low}. The resolution of this setup is (384,768). In Fig. \ref{fig:earthhigherst} we present the results for the TVA model and as we increase the Stokes number the evolution of the gas around the planet in the TVA model is in good agreement with the Two Fluid model. We again see the ring-like feature close to corotation, which is absent in the two-fluid models.

When increasing the Stokes number above 0.05 to 0.1 and 0.2, we see the features around the planet after 500 orbits being to spread out to further radii from the planet. At first it may seem to be a resolution problem as the Earth mass planet has small scale interactions with the dust around its orbit, however upon doubling the resolution for the case of St $= 0.2$ we find the same deviation of the dust evolution around the planet. This is still in line with the limits due to the terminal velocity approximation being applied where the Stokes number should be St $\ll 1$ for the model to be accurate. 

The presence of these features is clear in the different cases of the Stokes number. When considering the lower Stokes numbers, the features in the horseshoe region are very narrow. As we increase the Stokes number, these features become more prominent and spread out to a further radius from the planet's position. Based on our three cases, we plot the dust density differences between the Two Fluid model and the TVA model. For the gas, in all three cases, the density differences between the two models are at most $0.3\%$, hence are not included in the figure. In Fig. \ref{fig:densdiff}, we have set the percentage limits of dust density difference to be $\pm5\%$ through the scale itself. As we increase the Stokes number above 0.05 to the cases of 0.1 and 0.2 we see that the region where the difference between the two models is outside $\pm5\%$ extends out to further radii from the planet and is more prominent in the horseshoe region of the planet.

In summary, for low-mass planets the TVA and two-fluid models agree very well on the wave structure in the disc for Stokes numbers up to ${\rm St}=0.1$. TVA shows an artificial dust enhancement in the inner disc, which is due to the source term in the internal energy equation and can be remedied by solving for the entropy \citep{mcnally19a}. Additionally, TVA shows a dust ring at corotation that is absent from two fluid models (see also section \ref{sec:lobes}).

\subsection{Larger Planet Masses}\label{large}

Having shown in the previous sections the agreements and limitations of the TVA model to the Two Fluid model, we present the results of having a higher planet mass embedded in the disc than an Earth sized mass. In the following results we use a Stokes number of St $=0.01$ and resolution of (768,1536), having shown in previous sections there is a good agreement between the two models up to an Earth sized mass embedded. From \cite{lovascio19}, we know that the TVA model can break down around shocks as we explore how a higher planet mass can affect the evolution of the protoplanetary disc being modelled as a single fluid.

\begin{figure}
\includegraphics[scale=0.35]{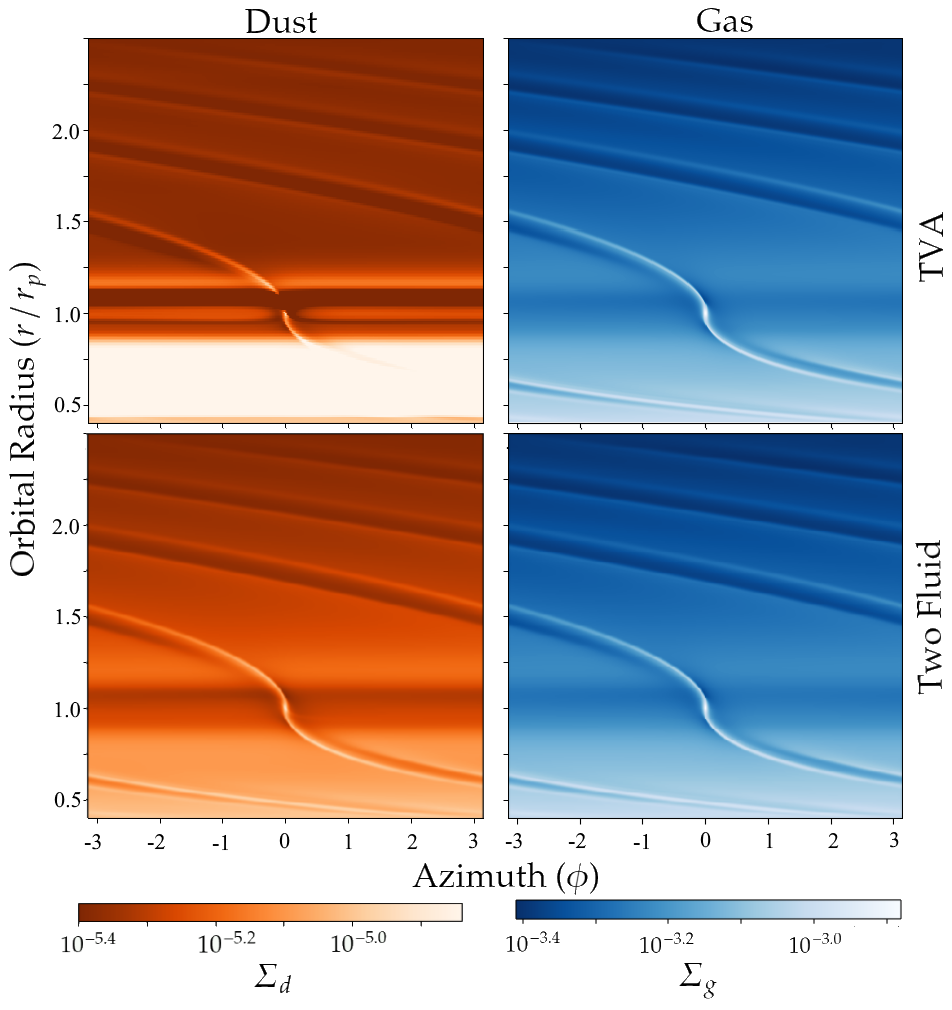}
\caption{Dust (left column) and gas (right column) surface density plots for the TVA (top row) and Two Fluid (bottom row) models. The disc is embedded with a Neptune mass planet at 1 $r_{\rm p}$. The time period $T_p = 100$ orbits, Stokes number St$=0.01$ and dust fraction $f_{\rm d}=0.01$.} 
\label{fig:neptunecomb}
\end{figure}

\begin{figure}
\includegraphics[scale=0.35]{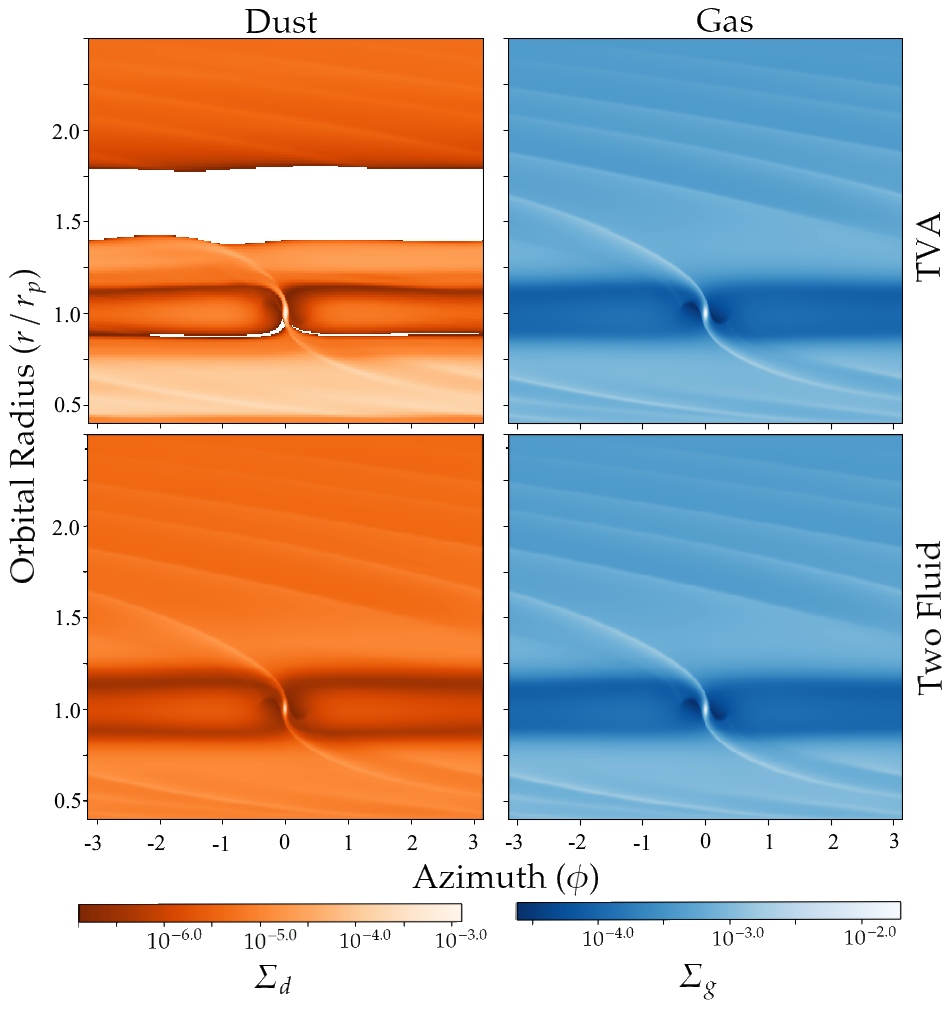}
\caption{Dust (left column) and gas (right column) surface density plots for the TVA (top row) and Two Fluid (bottom row) models. The disc is embedded with a Jupiter mass planet at 1 $r_{\rm p}$. The time period $T_p = 100$ orbits, Stokes number St$=0.01$ and dust fraction $f_{\rm d}=0.01$.} 
\label{fig:jupitercomb}
\end{figure}

\begin{figure}
\centering
\includegraphics[scale=0.35]{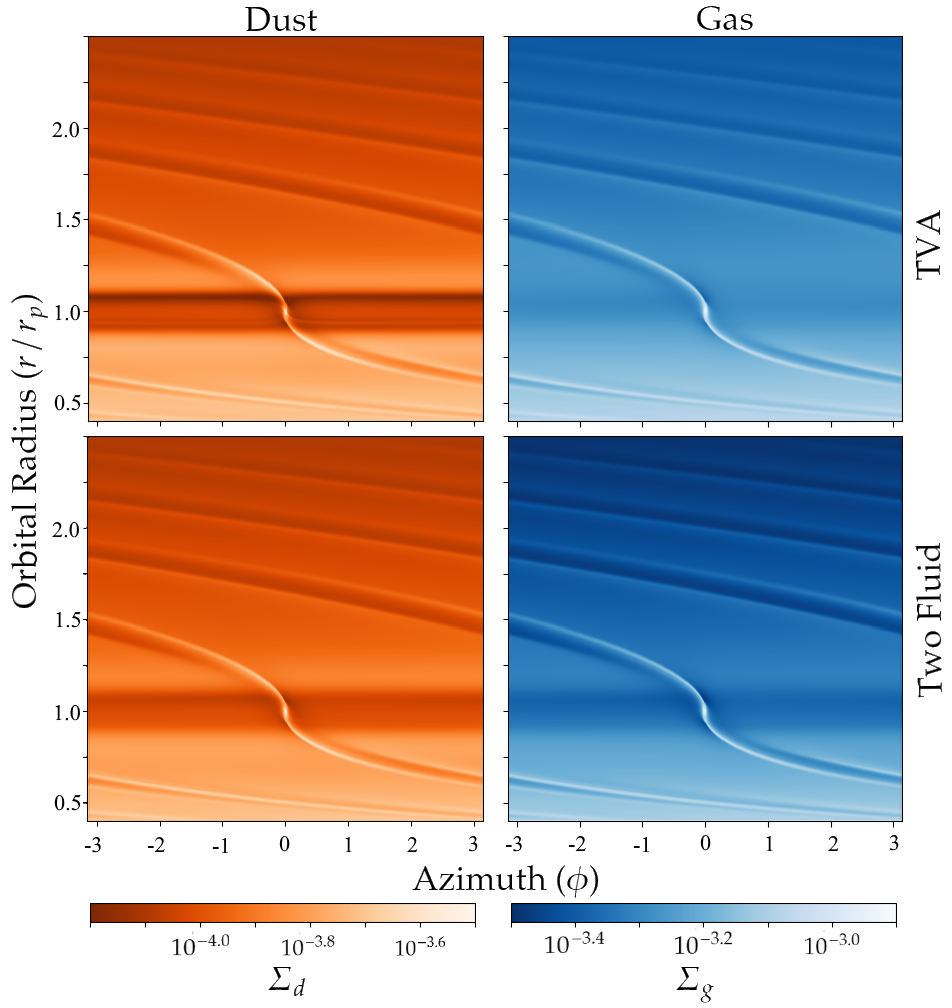}
\caption{Dust (left column) and gas (right column) surface density plots for the TVA (top row) and Two Fluid (bottom row) models. The disc is embedded with a Neptune mass planet at 1 $r_{\rm p}$. The time period $T_p = 100$ orbits, Stokes number St$=0.01$ and dust fraction $f_{\rm d}=0.2$.} 
\label{fig:neptunecomb02}
\end{figure}

\begin{figure}
\centering
\includegraphics[scale=0.35]{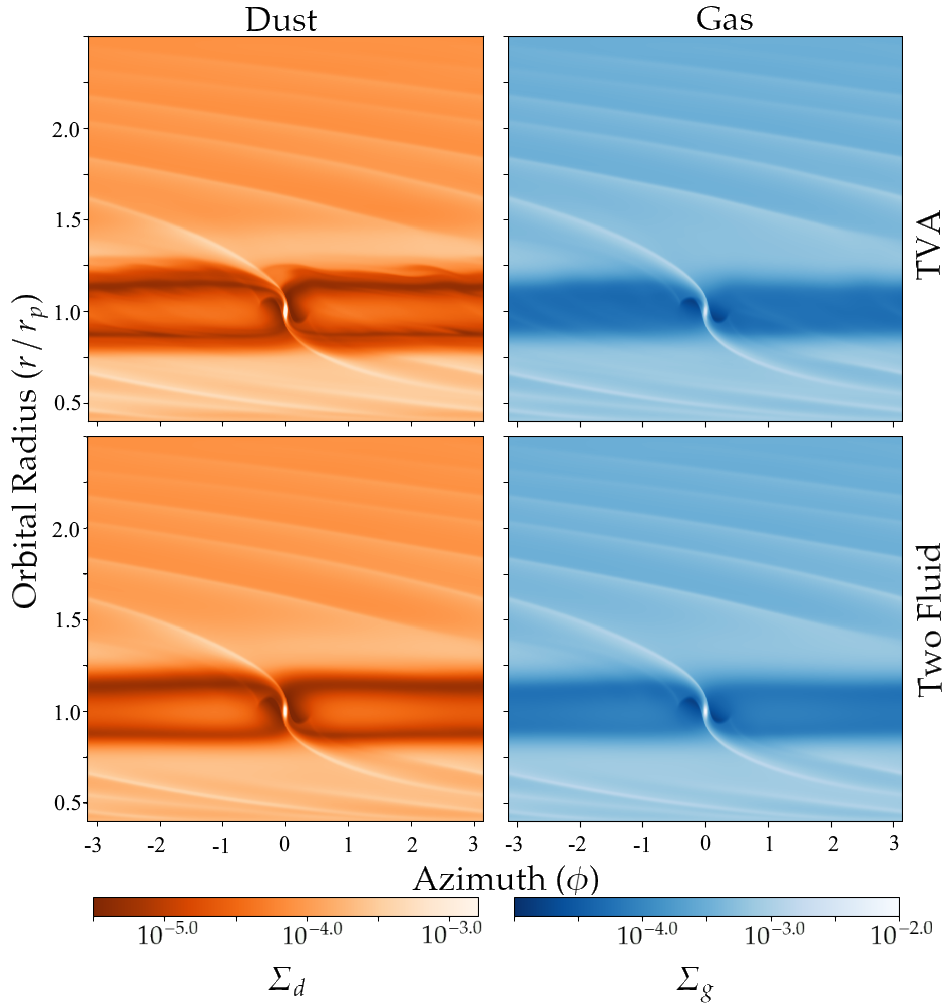}
\caption{Dust (left column) and gas (right column) surface density plots for the TVA (top row) and Two Fluid (bottom row) models. The disc is embedded with a Jupiter mass planet at 1 $r_{\rm p}$. The time period $T_p = 100$ orbits, Stokes number St$=0.01$ and dust fraction $f_{\rm d}=0.2$.} 
\label{fig:jupitercomb02}
\end{figure}

\begin{figure*}
\centering
\includegraphics[scale=0.4]{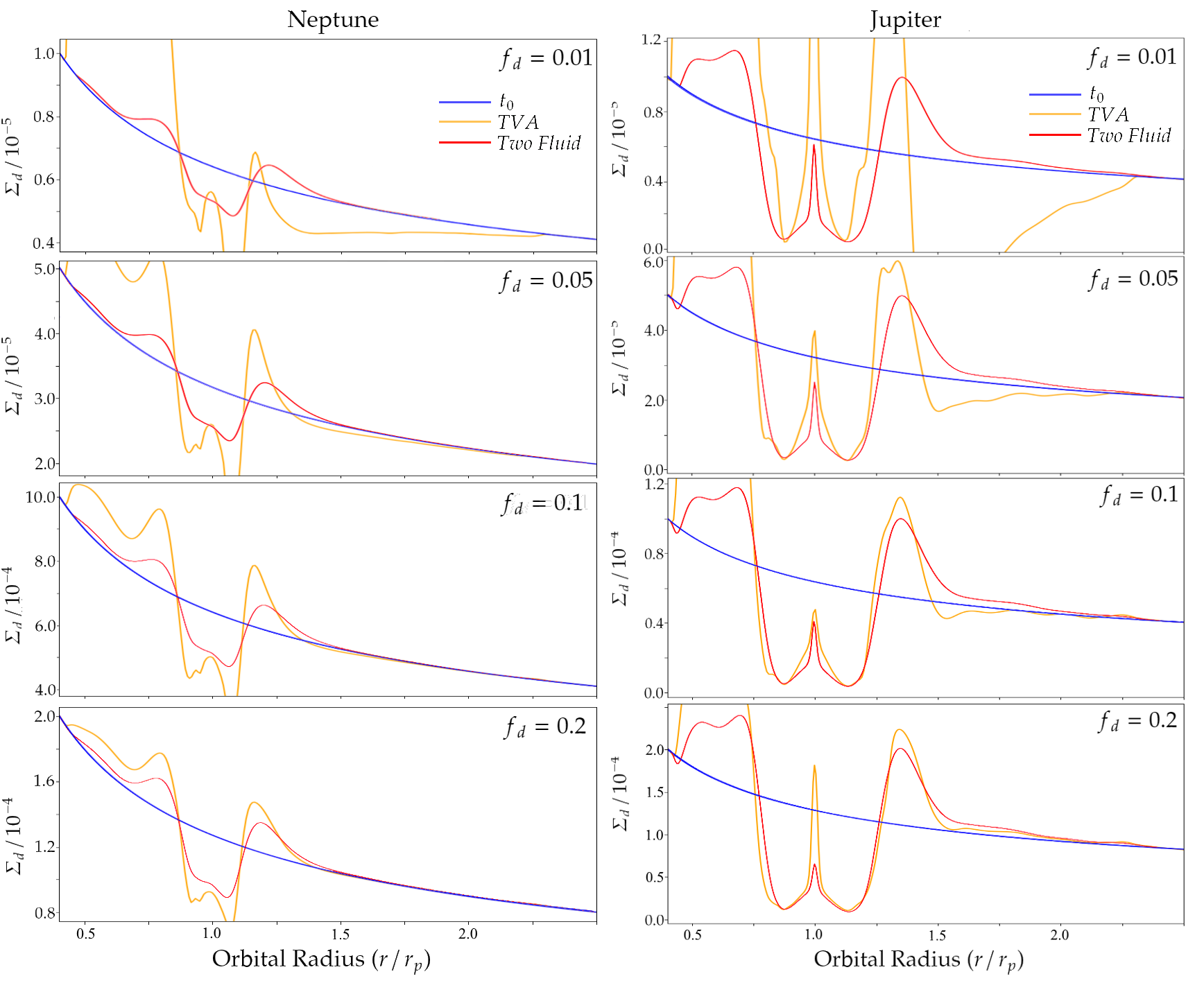}
\caption{Azimuthally averaged dust surface density for increasing dust fractions, $f_d = 0.01, 0.05, 0.1, 0.2$ with a Neptune (left column) and Jupiter (right column) mass planet embedded at initial time (blue line) and after 100 orbits for TVA (yellow line) and Two Fluid (red line) models}
\label{fig:combodrifts}
\end{figure*}

In Fig. \ref{fig:neptunecomb}, we present the evolution of the dust and gas density with a Neptune mass planet, $10^{-4}M_{P}$, embedded after 100 orbits and in Fig. \ref{fig:jupitercomb}, the case of a Jupiter mass planet, $10^{-3}M_{P}$, embedded after the same time period. In the case of a Neptune mass planet embedded in the disc with a dust ratio, $f_{\rm d} = 0.01$, the spiral density waves produced by the planet through the disc are what is expected. However, we can see that after period of 100 orbits, there is a dust enhancement interior to the planet which we saw in the case of an Earth sized planet embedded with small dust fraction. The dust enhancement for the Neptune mass planet embedded disc is much larger than the Earth mass embedded case. Exterior to the planet, the evolution compared to the Two Fluid model is more comparative with an expected gap and pressure bump however the gap carved is much deeper. For the Jupiter mass case in Fig. \ref{fig:jupitercomb}, the evolution of the dust breaks down exterior to the planet with the dust density dropping below zero while the interior has again an increasingly larger dust enhancement. For both cases we see that the gas evolution is accurate to the theory with the Jupiter mass planet carving a deeper gap in the disc.

The dust enhancement interior to the planet appears during the early stages of the evolution of the disc and this initial dust enhancement causes a significant change to the evolution of the dust over time. We find that the relative size of the initial dust enhancement is independent of the initial dust ratio and is dependent on the mass of the planet embedded. As we consider a higher dust fraction, the contribution from the initial dust enhancement does not affect the dust evolution as greatly, as we saw from Fig. \ref{fig:earthcomps005} where with a higher dust fraction, $f_{\rm d} = 0.05$ the dust enhancement interior to the planet is not observed after 500 orbits compared to Fig. \ref{fig:earthcomps001}. We explore this further with the larger planets where a higher dust fraction is modelled and the results of the azimuthally averaged surface density for the dust. 

In Fig. \ref{fig:neptunecomb02} and \ref{fig:jupitercomb02} we present the density plots for the $f_d = 0.2$ case for the Neptune and Jupiter embedded disc respectively to see how the structures differ between the two models. Compared to the lower dust fraction case in Fig. \ref{fig:neptunecomb} and \ref{fig:jupitercomb}, we see that the dust evolution for the TVA model resembles the Two Fluid model with a difference of a larger dust buildup at the pressure bumps on the edge of the gaps carved by both planets as indicated by the azimuthally averaged data. As shown, the higher initial dust density is largely unaffected by the initial dust enhancement caused by the presence of the planet, thereby following the expected evolution of a gap being carved and dust rings forming interior and exterior yet still differing from the Two Fluid model in terms of the magnitudes. The TVA model in Fig. \ref{fig:jupitercomb02} shows hints of an unstable gap edge in the dust density with swirly structures around $r/r_p=1.25$, while these features are absent in the two fluid model.

In Fig. \ref{fig:combodrifts}, we present the azimuthally averaged dust surface density for dust fraction $f_{\rm d}\in [0.01,0.05,0.1,0.2]$ in the Neptune and Jupiter embedded case for both models. We see in the Neptune mass planet case that as the dust fraction is increased, the effect of the initial dust enhancement is lessened, relative to the dust density as it dampens the bump more smoothly from an earlier time. The same situation applies for the Jupiter mass planet case with the dust evolution being less affected by the initial dust enhancement. The evolution of the dust in the TVA model differs from the Two Fluid model in terms of the size of the dust rings interior and exterior to the planet compared to the earlier tolerable limit for Earth mass planet embedded disc. When comparing the averaged dust surface density for both the models we see that the higher dust fraction cases for the LITVA model creates a more pronounced outer and inner ring. The differences are therefore likely to be due to shocks not being treated correctly in TVA. 

Regarding the evolution of protoplanetary discs with a planet embedded, we conclude that currently the TVA model modified in FARGO3D is of limited use for evolving the dust and gas together as a single fluid for a larger planet mass than Earth sized. It induces a dust enhancement relative to the mass of the planet embedded, which affects the evolution of the dust over time drastically. With an increasing higher dust fraction, the model produced a gap and dust rings however do not completely replicate the Two Fluid results to a tolerable limit even with a Stokes number $\ll 1$.

\subsection{Computational Cost}\label{cost}

Initially we tested the performance of the TVA model in terms of the real time runs of an Earth embedded disc on four cores. For these runs, we used a resolution of (384,768) and a simulation time of, $T_{\rm p} = 500$ orbits. The dust fraction was set to 1\% and Stokes numbers varied between $10^{-1}$ to $10^{-6}$. The parameters were chosen for the agreement between the two models in terms of the gas and dust evolution shown in the previous sections. Larger planet mass runs were not tested for their computational time due to the differences in the evolution between the TVA model and Two Fluid model. However, different dust fractions were tested with the Earth mass planet embedded, but not included in the results as there were negligible differences made in real run times compared to Table \ref{table:tvatime}. We present the results for the run times of these setups in Table \ref{table:tvatime} where the protoplanetary disc was evolved with different Stokes numbers, timing three runs for each value and averaging between the three. We can see that for different Stokes numbers the real run times did not differ much between runs for the TVA model. Similarly we timed Two Fluid runs as well to see if there would be any differences when changing the Stokes number but they were also consistent from $10^{-1}$ to $10^{-6}$. The similarity between the run times for each model would be due to the implicit methods of numerical modelling implemented to evolve the dust and gas in FARGO3D. 

We note that a lower Stokes number for the Two Fluid model of $10^{-9}$ was tested however the evolution of the disc broke down which was due to the parameter having order close to machine precision therefore ruling out it's ability for evolution of the disc under a perfect coupling regime without modification. For the TVA model, the Stokes number can be set down to 0 for perfect coupling and achieves it's expected evolution of the dust following the gas exactly, which is stable over the 500 orbits.

\begin{table} 
\centering
{TVA}
\bigskip
    \begin{tabular}{lllll} \hline
St    & $t_1$  & $t_2$  & $t_3$  & Average Time  \\ \hline
$10^{-1}$ & 56m 12.164s & 55m 30.531s & 55m 36.012s & 55m 46.236s \\ 
$10^{-2}$ & 56m 31.401s & 55m 39.483s & 55m 50.323s & 56m 0.402s \\ 
$10^{-3}$ & 56m 34.536s & 56m 41.645s & 57m 43.612s & 56m 59.931s \\ 
$10^{-4}$ & 56m 5.854s & 56m 12.870s & 56m 2.712s & 56m 7.145s \\ 
$10^{-6}$ & 58m 23.445s & 58m 34.139s & 58m 36.214s & 58m 31.266s \\  \bottomrule
\end{tabular}
    \caption{TVA real time runs for an Earth embedded disc with $f_d = 0.01$ over 500 orbits.}
    \label{table:tvatime}
\end{table}

\begin{figure}
\centering
\includegraphics[scale=0.34]{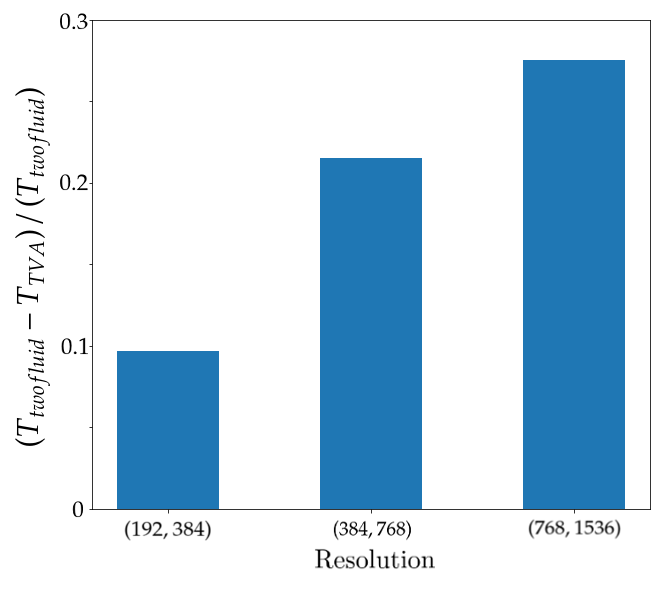}
\caption{Real time runs difference between the Two Fluid model and TVA model for resolutions $\in (192,384),(384,768)$ and $(768.1536)$.} 
\label{fig:resolution}
\end{figure}

\begin{figure}
\centering
\includegraphics[scale=0.34]{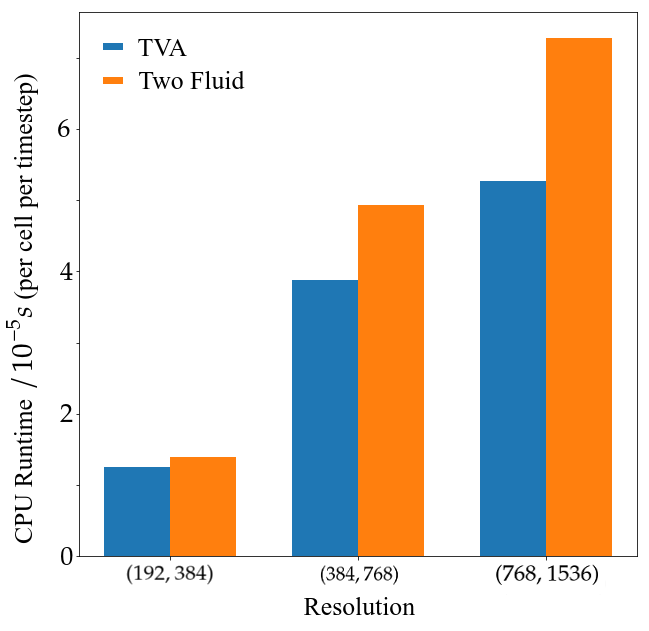}
\caption{CPU runtime per cell, per timestep for the Two Fluid model and TVA model for resolutions $\in (192,384),(384,768)$ and $(768.1536)$.} 
\label{fig:percell}
\end{figure}

When comparing the average times between the two models, the TVA model completed the 500 orbits faster and we explore how the computational cost varies with resolution. We set up three different resolutions of $(192,384),(384,768)$ and $(768,1536)$ and ran the simulations for 200, 20 and 5 orbits respectively on four cores. For each resolution, we ran the setup five times and took an average before calculating the percentage differences between the two models. In Fig. \ref{fig:resolution} we present the results for the difference between the two models in terms of real run times and in Fig. \ref{fig:percell} we present the CPU runtimes for the two models per cell and per timestep. We see that for a resolution of $(192,384)$ we get a percentage reduction of $9.67\%$ in real time computational cost. When doubling the resolution to $(384,768)$ the percentage reduction we found was $21.54\%$ and further doubling the resolution to $(768,1536)$, a reduction in real time computational cost of $27.52\%$. More resolution tests would have to be considered to find an accurate relation of how the computational cost reduction scales with resolution, however we show that for a high resolution Earth mass planet embedded disc, the TVA model saves a considerable amount of computational time compared to the Two Fluid model. 

Since we use a semi-implicit method (\citealt{meyer14}) for solving our cooling term there is a "maximum stable super-step" allowed for this RKL2 scheme which gives a time-step constraint on our simulations, $\Delta t_{\rm max} = \Delta t_{\rm expl} (s^2+s-2)/4 $ with $\Delta t_{\rm expl}$ being the maximum stable explicit time-step for parabolic terms and $s$ as the number of Runge-Kutta stages. For a much larger resolution and/or diffusion coefficient there would be a smaller maximum super-step allowed, contributing to a higher computational cost. We note however that our diffusion coefficient for the cooling term is $D = c_s^2 t_s f_d$ and given the typical values we have used for our simulations, the magnitude of the coefficient is still very small compared to the grid cell sizes we have used and tested in this paper. 

Altogether, we show that in the regime of strong drag, the TVA model is much more ideal when simulating evolutions of the disc over long simulation times as it is much less taxing in terms of its computational cost while producing an accurate evolution of the gas and dust evolution in an Earth embedded disc. This advantage is furthered when considering simulations with a higher spatial resolution which takes an increasing amount of computational cost.

\section{Summary and Discussion} \label{discuss}

\subsection{Agreements}

In this paper we have demonstrated that the dust and gas under the strong drag regime can be effectively modelled as a single fluid over a long time period. Our results show that the evolution for an axisymmetric disc and an Earth planet mass embedded are comparative with a Two Fluid model while reducing the computational cost from the difficulty of accounting for the interactions between well coupled dust and gas. Interactions such as the drag that the gas applies to the dust and the back-reaction from the dust are accounted for through the relative velocity evolution which is simplified under the small stopping time (Equation \ref{tvamixrel}). This provides a useful basis for simulations of protoplanetary discs and possible substructures whereby the dust and gas could be tightly coupled.

\subsection{Origin of the corotation dust ring}
\label{sec:lobes}

For the Earth mass planet embedded case, a dust feature was present in the horseshoe region of the planet in the TVA model and in \ref{differences}, we compared the differences and possible implications of this dust feature for the Stokes number $St = 0.01$ case. It is either an artifact of the TVA model, or a genuine feature that remains unresolved in two fluid calculations. Here we hypothesize that this ring may actually be a feature that is seen in a previous study of 3D simulations of planet embedded gaseous discs with thermal diffusion (\citealt{lega2014}). The feature found was the formation of an asymmetric cold finger of gas extending from the planet in the early evolution of the disc after 10 orbits. This cold finger is caused by thermal diffusion. In an adiabatic disc, the planet creates a hot atmosphere. Adding thermal diffusion means that this heat will spread, and this spread will be asymmetric if the planet is not exactly located at corotation. This will in general be the case in a gas disc that has a radial pressure gradient. Depending on whether the planet is luminous enough, the resulting finger will be cold (no or low luminosity) or hot \citep[high enough luminosity][]{2017MNRAS.472.4204M}. 

In the thermodynamic view of a dusty gas \citep{lin17}, a temperature increase corresponds to a decrease in dust fraction. A 'luminous' planet would then be a planet that accretes dust but not gas. Since we do not consider accretion on the planet, we are dealing with a non-luminous planet and therefore the 'cold finger' \citep{lega2014}, or a dust-rich finger, which is what is observed. The 'thermal diffusion' responsible would be the cooling term within TVA. It should be noted that this cooling term has a different form compared to thermal diffusion, not least because it acts on the pressure rather than the temperature.

These thermal lobes (hot or cold), are notoriously difficult to resolve. For thermal diffusivity $\chi$, the size of the lobes is $\sim \sqrt{\chi/\Omega_p} \ll H$ \citep{2017MNRAS.472.4204M}. In the thermodynamic picture of TVA, we have a diffusion coefficient $\sim c_s^2t_s$, suggesting again scales much smaller than $H$. 

We ran simulations with a higher resolution of (1200,3600), higher dust fraction and Stokes number to try and resolve the cold finger and associated thermal lobes. From preliminary tests we see that the width of the thermal lobes and angle of the cold finger roughly matches results from \cite{lega2014}. When the cooling term is turned off, the dust artifact disappears which indicates that the artifact is due to our "thermal diffusion". For both Two Fluid and LITVA regimes, the requirements to resolve the phenomenon are very high and in the Two Fluid model, similar setups show hints of thermal lobes but are more diffuse as the resolution requirement is more severe in the Two Fluid model than LITVA. The resolution requirement given in \cite{laibe12} is $\Delta x < v_{\rm gas} t_s$, making sure that the length scale over which the gas adjusts to the dust is resolved. Since in the corotation region the gas velocity $v_{\rm gas} \ll c_s$, and the lobes are located very close to corotation, it is likely that in order to see the thermal lobes in two fluid calculations resolutions of 1000s of cells per scaleheight are necessary. However, as was pointed out in \cite{2018A&A...617A.117R}, the resolution criterion of \cite{laibe12} appears to specific to SPH, as they show PLUTO does not suffer from it.  

We stress that more work will be necessary to confirm whether this is the origin of the dust feature as our diffusion term is different to thermal diffusion and radiative diffusion as considered in \cite{lega2014} and \cite{2017MNRAS.472.4204M}. Additionally, in order to replicate this phenomenon in 2D simulations, \cite{lega2014} had to use a much smaller smoothing length for the planet's potential than the standard smoothing length of $0.6H$. In our case, the thermal lobes show up for a standard value of the smoothing length. It is worth noting that even if thermal lobes are related to the corotation dust ring that we observe, it is possible that the TVA artificially enhances this.

If these are indeed dusty versions of the thermal lobes, they can have important consequences for the torque on the planet and subsequent migration. In terms of the torque exerted on the planet by the dust flow around the planet, previous research by \cite{BenLlam2018} has shown that asymmetries in the dust density distribution can play a crucial role in determining the net torque on the planet. Thermal lobes would come on top of that. Future work should focus on whether an exact correspondence exists between thermal lobes and the dust feature we observe, and what the sign and magnitude of the torque is.

\subsection{Limitations}

\subsubsection{Dust Buildup}

Our simulations have shown that for a larger planet mass, the LITVA model does not replicate the same results from a Two Fluid model. For a Neptune sized planet and above the evolution of the disc differed greatly due to a dust buildup interior to the planet and enhanced dust rings. Currently, we rule out the use of the LITVA model in FARGO3D in this regime due a number of issues that have been demonstrated in this paper and theories from previous studies. This includes the dust enhancement which we hypothesised was due to the form of the pressure evolution equation (Equation \ref{pressevo}) when evolved in place of the energy equation in FARGO3D, the source term can adversely affect how the model evolves via a dust enhancement as the density is dictated by the pressure (Equation \ref{dustfracpressure}). As mentioned, a workaround was implemented by \cite{mcnally19a} through a change in the form of the evolution equation, which prevented a similar situation.

\subsubsection{Shocks}

In \cite{lovascio19}, the TVA model was shown to break down around shocks and we believe that the scaling of the dust enhancement with respect to the planet mass embedded would be tied in to this previous result as the initial dust density is greatly affected by the early evolution of the disc as a large planet interacts with it through the launching of density waves. Solutions mentioned (see \cite{lovascio19}, Section 4.2) were unsuitable as it required limiting resolution or introducing a higher shock viscosity when resolving the shock. The difference between LITVA and the Two Fluid model, notably the strength of the dust rings outside the orbit of the planet, can be attributed to the effect of shocks. Note that these are features that are important for interpreting observations, so that care must be taken when using one fluid models to interpret for example ALMA observations. Note that the severity of the discrepancy between TVA and two fluid probably depends on many parameters, such as the overall dust fraction, but also on viscosity and whether simulations are carried out in 2D or 3D. 

\subsection{Varying Parameters}

\subsubsection{Kinematic Viscosity}

With the corotation features appearing in the LITVA model, we explored the possible differences with a lower and higher kinematic viscosity for the Earth mass planet embedded case. We changed the kinematic viscosity in the setup from $\nu = 10^{-5}$ to $\nu \in [10^{-4},10^{-6}]$, keeping the dust fraction and Stokes number the same as previous, both $0.01$. For the higher kinematic viscosity of $10^{-4}$, we find that the two models differ by roughly the same amount as in the $10^{-5}$ case where the maximum percentage differences in dust density between the two models are an enhancement of $+5.18\%$ and a reduction of $-2.10\%$ over the whole domain after 2000 orbits. From 500 to 2000 orbits this value changes by around $\pm0.01\%$ which indicates a stable configuration as the increase in kinematic viscosity lessens the effect of the artifact in the LITVA model over time compared to the lower viscosity case. When lowering the kinematic viscosity further to $10^{-6}$ we observe a maximum percentage difference in the dust density between the two models of $+5.30\%$ and $-4.20\%$ after 2000 orbits. From 500 to 2000 orbits in this setup, the maximum percentage reduction grew from $-3.17\%$ to $-4.20\%$ which shows that more care should be taken when using the LITVA model for setups with lower viscosity and longer integration times. The region where this reduction difference occurs is located behind the orbit of the planet in the corotation dust ring.

\subsubsection{Aspect Ratio}

The other parameter we explored was the aspect ratio of the disc. We ran simulations with lower and higher disc aspect ratios, compared to the standard $0.05$, of $H/r \in [0.01,0.1]$ with an Earth mass planet embedded for 2000 orbits. This corresponds to a planet mass of $3.0$ and $3\times10^{-3} M_{th}$ respectively where the thermal mass (\citealt{2001ApJ...552..793G}) is given as $M_{th} = (H_p/r_p)^3M_P$ with $H_p$ as the disc scale height at the planet's location. With the presence of the dust ring at corotation, we expected that a thinner aspect ratio disc would be more easily perturbed during its evolution and we see that with a disc aspect ratio of $0.01$, the LITVA model drastically diverges from the Two Fluid model in the horseshoe region where a gap is created that spreads from planets location radially. With a higher aspect ratio of $0.1$ the two models agree much more closely as the dust ring is smoothed out more and affects the disc less with maximum dust density percentage differences of $+2.13\%$ in a few cells around the planet and $-0.64\%$ in the horseshoe region of the planet.

When the disc aspect ratio is unchanged from $[0.01,0.1]$ but a Neptune planet mass is embedded, corresponding to $100$ and $0.1 M_{th}$ we find that for the lower disc aspect ratio, the Neptune sized planet carves out a gap where the evolution of the dust breaks down with the dust density dropping below zero creating a larger discrepancy between the two models than the disc aspect ratio of 0.05 case. For the higher disc aspect ratio the LITVA model agrees closely with the Two Fluid model in all regions apart from the corotation region behind the planet as a shallow cavity is created in the middle of the asymmetric horseshoe leg. This contrasts to the results in Fig. \ref{fig:neptunecomb} with a disc aspect ratio of $0.05$, corresponding to a planet mass of $0.8 M_{th}$ where the LITVA model did not agree well with the Two Fluid model.

We then ran simulations with a Neptune mass planet with disc aspect ratios of $H/r \in [0.03218... , 0.3218...]$ for 200 orbits to obtain a planet mass of $3$ and $3\times10^{-3} M_{th}$ respectively to see if the LITVA model produced the same agreements and disagreements with the Two Fluid model as the Earth mass planet embedded disc with the same thermal mass. We find that with the lower disc aspect ratio, the results diverge from the Two Fluid model with a dust enhancement interior to the planet and gaps near the planet with the dust density dropping below zero. The higher aspect ratio simulations resulted in the two models agreeing closely in all regions apart from one cell width from the radial boundaries. In the one cell width from the inner and outer boundary, the dust density percentage difference was larger than $\pm5\%$ between the two models and everywhere else never exceeded $+0.65\%$ and $-0.23\%$.

With these results and previous ones in subsection \ref{large}, we show that increasing the planet mass while keeping the disc aspect ratio constant leads to disagreements between the two models as does keeping the planet mass constant and decreasing the disc aspect ratio. Therefore it would seem that the thermal mass can play an important role in how well the LITVA model agrees with Two Fluid since keeping the same thermal mass for a different planet mass and disc aspect ratio produces the same comparison between the two models. Since the thermal mass is a measure of nonlinearity in the flow and therefore of shocks appearing close to the planet \citep{1996ApJS..105..181K}, it is understandable for planets more massive than a thermal mass in the terminal velocity approximation.

\subsubsection{Softening Length}

We tested the evolution with a smaller softening length using a coefficient of $0.2 R_H$ where $R_H$ is the Hill sphere radius of the planet given by, $R_H = r_p(q/3)^{1/3}$ with $q$ as the planet-to-star mass ratio. We expected that a smaller softening length would create larger discrepancies between the two models as the interaction between the planet and the single fluid mixture in LITVA would be stronger closer to the planet.

We tested this setup with an Earth, Neptune and Jupiter mass planet and found that for the Earth mass planet embedded disc, the dust enhancement interior to the planet greatly increased by a factor of 6 rendering it incomparable to the Two Fluid model. While such a small softening is usually thought inappropriate for a low-mass planet, we note that it leads to stronger density waves, which may be responsible for the stronger discrepancy in the inner disc. 

For the Neptune and Jupiter mass planet embedded discs we found discrepancies in the dust density and radial velocity profiles when comparing the two softening lengths for the LITVA model. The smaller softening length created sharper gradient changes in the velocity profile and a large dust density buildup around the planet's location for both cases. Another discrepancy was the creation of gaps and rings where the dust density reached below zero, similar to our previous results of a Jupiter mass planet embedded disc in Fig. \ref{fig:jupitercomb}.

\subsection{Use of TVA and Outlooks}

Although the use of the approximation in a protoplanetary disc with a planet embedded has mostly been limited to specific setups, the results from previous studies have shown good agreement in replicating planet disc interactions. In \cite{chen18}, interactions such as dust trapping by pressure bumps, dust settling and streaming instability have been reproduced in the TVA model when evolving a dusty disc and in \cite{ballabio18} the approximation was implemented into PHANTOM SPH to simulate a disc in 3D with large planets embedded. More work however will be needed to continue the evaluation of the use of TVA in modelling the evolution of protoplanetary discs for a variety of situations and setups. From our results, when starting dusty discs with a pressure gradient, we have shown that for large planet masses embedded into the disc, caution should be taken when implementing the model as evolution around the planet diverges from expectations. We see that for up to Earth mass planet embedded, the TVA model provides a good advantage in computational cost over a Two Fluid for evolving the dust and gas as a single fluid with good agreement between the two models. Therefore with a tightly coupled gas and dust disc with a low mass planet the LITVA model would be reasonable to use especially with long evolution times. This would correlate with up to an Earth mass planet and a Stokes number of up to 0.05 for a planet embedded dusty disc. In addition, LITVA has a clear advantage over two fluid in terms of the required resolution, especially if features close to the planet are important (see section \ref{sec:lobes}).

A general caveat of using the terminal velocity approximation is when the dust starts to decouple from the gas in realistic situations. This could happen due to a variety of reasons, for example, in dust rich areas of a protoplanetary disc where planetesimals could form. Therefore caution should be taken when using the TVA model to model protoplanetary discs compared to the two fluid approach. Optimally, a combination of the two models would be ideal if considering global disc simulations where the strength in the coupling between the dust and gas is continuously changing. 

Additionally, comparisons between the LITVA model and Two Fluid model would be interesting when it comes to running 3D simulations in Eulerian methods. The evolution of the dust and gas as a single fluid would require additional consideration due to the reduction of the dust fraction with height in the disc and the increase in stopping times. The reduction in computational cost when modelling the dust and gas as a single fluid would also be of interest as we would expect a larger relative reduction in computational costs in 3D as we would be reducing 8 evolution equations down to 5 rather than 6 to 4 in 2D.

\section{Conclusions} \label{conclusions}

We performed 2D simulations comparing the evolution of the protoplanetary disc between the Locally Isothermal Terminal Velocity Approximation model and Two Fluid model. Implementation of the LITVA model in FARGO3D has shown that the approach of modelling dust and gas as a tightly coupled single fluid can be used to study the evolution of a protoplanetary disc with an embedded planet less massive than the Earth, closely matching the evolution of the Two Fluid model. The ability to model the dust and gas together as a single fluid and then recover their respective densities and velocities is important for tightly coupled models which implement explicit methods to evolve the dust and gas separately since they require smaller time steps. 

\begin{itemize}
    \item {The computational cost is much lower, saving up to $27.52\%$ computational time for higher resolution runs in the LITVA model compared to the Two Fluid approach even for implicit methods where there is no timestep constraint unlike explicit methods which require smaller timesteps for stability.}
    
    \item {We have found good agreement between LITVA and two fluid for Earth-mass planets, except for the corotation density feature that we tentatively attribute to dusty 'thermal lobes' \citep{2017MNRAS.472.4204M}.}
    
    \item {For larger planet masses, the evolution of the dust does not match Two Fluid simulations even in the strong drag regime due to a combination of an artificial "cooling" of the inner disc and stronger shocks introduced by larger planets. This has important implications for interpreting for example ALMA observations using one fluid models.}
\end{itemize}

\section*{Acknowledgements}

KC is funded by an STFC studentship. SJP is funded by a Royal Society University Research Fellowship.
\addcontentsline{toc}{section}{Acknowledgements}

\section*{Data Availability}

The base code of FARGO3D is publicly available at \url{https://bitbucket.org/fargo3d/public.git}. The modified version used to perform the calculations in this work will be shared on reasonable request to the corresponding author.




\bibliographystyle{mnras}
\bibliography{example} 

\begin{thebibliography}{}
\makeatletter
\relax
\def\mn@urlcharsother{\let\do\@makeother \do\$\do\&\do\#\do\^\do\_\do\%\do\~}
\def\mn@doi{\begingroup\mn@urlcharsother \@ifnextchar [ {\mn@doi@}
  {\mn@doi@[]}}
\def\mn@doi@[#1]#2{\def\@tempa{#1}\ifx\@tempa\@empty \href
  {http://dx.doi.org/#2} {doi:#2}\else \href {http://dx.doi.org/#2} {#1}\fi
  \endgroup}
\def\mn@eprint#1#2{\mn@eprint@#1:#2::\@nil}
\def\mn@eprint@arXiv#1{\href {http://arxiv.org/abs/#1} {{\tt arXiv:#1}}}
\def\mn@eprint@dblp#1{\href {http://dblp.uni-trier.de/rec/bibtex/#1.xml}
  {dblp:#1}}
\def\mn@eprint@#1:#2:#3:#4\@nil{\def\@tempa {#1}\def\@tempb {#2}\def\@tempc
  {#3}\ifx \@tempc \@empty \let \@tempc \@tempb \let \@tempb \@tempa \fi \ifx
  \@tempb \@empty \def\@tempb {arXiv}\fi \@ifundefined
  {mn@eprint@\@tempb}{\@tempb:\@tempc}{\expandafter \expandafter \csname
  mn@eprint@\@tempb\endcsname \expandafter{\@tempc}}}

\bibitem[\protect\citeauthoryear{{ALMA Partnership} et~al.,}{{ALMA Partnership}
  et~al.}{2015}]{alma15}
{ALMA Partnership} et~al., 2015, \mn@doi [\apjl] {10.1088/2041-8205/808/1/L3},
  \href {https://ui.adsabs.harvard.edu/abs/2015ApJ...808L...3A} {808, L3}

\bibitem[\protect\citeauthoryear{{Andrews} et~al.,}{{Andrews}
  et~al.}{2016}]{andrews16}
{Andrews} S.~M.,  et~al., 2016, \mn@doi [\apjl] {10.3847/2041-8205/820/2/L40},
  \href {https://ui.adsabs.harvard.edu/abs/2016ApJ...820L..40A} {820, L40}

\bibitem[\protect\citeauthoryear{{Andrews} et~al.,}{{Andrews}
  et~al.}{2018}]{andrews18}
{Andrews} S.~M.,  et~al., 2018, \mn@doi [\apjl] {10.3847/2041-8213/aaf741},
  \href {https://ui.adsabs.harvard.edu/abs/2018ApJ...869L..41A} {869, L41}

\bibitem[\protect\citeauthoryear{{Ayliffe}, {Laibe}, {Price}  \&
  {Bate}}{{Ayliffe} et~al.}{2012}]{ayliffe12}
{Ayliffe} B.~A.,  {Laibe} G.,  {Price} D.~J.,   {Bate} M.~R.,  2012, \mn@doi
  [\mnras] {10.1111/j.1365-2966.2012.20967.x}, \href
  {https://ui.adsabs.harvard.edu/abs/2012MNRAS.423.1450A} {423, 1450}

\bibitem[\protect\citeauthoryear{{Ballabio}, {Dipierro}, {Veronesi}, {Lodato},
  {Hutchison}, {Laibe}  \& {Price}}{{Ballabio} et~al.}{2018}]{ballabio18}
{Ballabio} G.,  {Dipierro} G.,  {Veronesi} B.,  {Lodato} G.,  {Hutchison} M.,
  {Laibe} G.,   {Price} D.~J.,  2018, \mn@doi [\mnras] {10.1093/mnras/sty642},
  \href {https://ui.adsabs.harvard.edu/abs/2018MNRAS.477.2766B} {477, 2766}

\bibitem[\protect\citeauthoryear{{Ballabio}, {Nealon}, {Alexander}, {Cuello},
  {Pinte}  \& {Price}}{{Ballabio} et~al.}{2021}]{ballabio21}
{Ballabio} G.,  {Nealon} R.,  {Alexander} R.~D.,  {Cuello} N.,  {Pinte} C.,
  {Price} D.~J.,  2021, \mn@doi [\mnras] {10.1093/mnras/stab922}, \href
  {https://ui.adsabs.harvard.edu/abs/2021MNRAS.504..888B} {504, 888}

\bibitem[\protect\citeauthoryear{{Baruteau}, {Fromang}, {Nelson}  \&
  {Masset}}{{Baruteau} et~al.}{2011}]{baruteau11}
{Baruteau} C.,  {Fromang} S.,  {Nelson} R.~P.,   {Masset} F.,  2011, \mn@doi
  [\aap] {10.1051/0004-6361/201117227}, \href
  {https://ui.adsabs.harvard.edu/abs/2011A&A...533A..84B} {533, A84}

\bibitem[\protect\citeauthoryear{{Baruteau}, {Wafflard-Fernandez}, {Le Gal},
  {Debras}, {Carmona}, {Fuente}  \& {Rivi{\`e}re-Marichalar}}{{Baruteau}
  et~al.}{2021}]{baruteau21}
{Baruteau} C.,  {Wafflard-Fernandez} G.,  {Le Gal} R.,  {Debras} F.,  {Carmona}
  A.,  {Fuente} A.,   {Rivi{\`e}re-Marichalar} P.,  2021, \mn@doi [\mnras]
  {10.1093/mnras/stab1045}, \href
  {https://ui.adsabs.harvard.edu/abs/2021MNRAS.505..359B} {505, 359}

\bibitem[\protect\citeauthoryear{{Ben{\'\i}tez-Llambay} \&
  {Masset}}{{Ben{\'\i}tez-Llambay} \& {Masset}}{2016}]{llambay16}
{Ben{\'\i}tez-Llambay} P.,  {Masset} F.~S.,  2016, \mn@doi [\apjs]
  {10.3847/0067-0049/223/1/11}, \href
  {https://ui.adsabs.harvard.edu/abs/2016ApJS..223...11B} {223, 11}

\bibitem[\protect\citeauthoryear{Ben{\'{\i}}tez-Llambay \&
  Pessah}{Ben{\'{\i}}tez-Llambay \& Pessah}{2018}]{BenLlam2018}
Ben{\'{\i}}tez-Llambay P.,  Pessah M.~E.,  2018, \mn@doi [The Astrophysical
  Journal] {10.3847/2041-8213/aab2ae}, 855, L28

\bibitem[\protect\citeauthoryear{{Ben{\'\i}tez-Llambay}, {Krapp}  \&
  {Pessah}}{{Ben{\'\i}tez-Llambay} et~al.}{2019}]{llambay19}
{Ben{\'\i}tez-Llambay} P.,  {Krapp} L.,   {Pessah} M.~E.,  2019, \mn@doi
  [\apjs] {10.3847/1538-4365/ab0a0e}, \href
  {https://ui.adsabs.harvard.edu/abs/2019ApJS..241...25B} {241, 25}

\bibitem[\protect\citeauthoryear{{Bohlin}, {Savage}  \& {Drake}}{{Bohlin}
  et~al.}{1978}]{bohlin78}
{Bohlin} R.~C.,  {Savage} B.~D.,   {Drake} J.~F.,  1978, \mn@doi [\apj]
  {10.1086/156357}, \href
  {https://ui.adsabs.harvard.edu/abs/1978ApJ...224..132B} {224, 132}

\bibitem[\protect\citeauthoryear{{Chen} \& {Lin}}{{Chen} \&
  {Lin}}{2018}]{chen18}
{Chen} J.-W.,  {Lin} M.-K.,  2018, \mn@doi [\mnras] {10.1093/mnras/sty1166},
  \href {https://ui.adsabs.harvard.edu/abs/2018MNRAS.478.2737C} {478, 2737}

\bibitem[\protect\citeauthoryear{{Cuello} et~al.,}{{Cuello}
  et~al.}{2019}]{cuello19}
{Cuello} N.,  et~al., 2019, \mn@doi [\mnras] {10.1093/mnras/sty3325}, \href
  {https://ui.adsabs.harvard.edu/abs/2019MNRAS.483.4114C} {483, 4114}

\bibitem[\protect\citeauthoryear{{Cuello} et~al.,}{{Cuello}
  et~al.}{2020}]{cuello20}
{Cuello} N.,  et~al., 2020, \mn@doi [\mnras] {10.1093/mnras/stz2938}, \href
  {https://ui.adsabs.harvard.edu/abs/2020MNRAS.491..504C} {491, 504}

\bibitem[\protect\citeauthoryear{{Dipierro}}{{Dipierro}}{2018}]{dipierro18a}
{Dipierro} G.,  2018, in Take a Closer Look. p.~69,
  \mn@doi{10.5281/zenodo.1488906}

\bibitem[\protect\citeauthoryear{{Dipierro}, {Price}, {Laibe}, {Hirsh},
  {Cerioli}  \& {Lodato}}{{Dipierro} et~al.}{2015}]{dipierro15}
{Dipierro} G.,  {Price} D.,  {Laibe} G.,  {Hirsh} K.,  {Cerioli} A.,   {Lodato}
  G.,  2015, \mn@doi [\mnras] {10.1093/mnrasl/slv105}, \href
  {https://ui.adsabs.harvard.edu/abs/2015MNRAS.453L..73D} {453, L73}

\bibitem[\protect\citeauthoryear{{Dipierro} et~al.,}{{Dipierro}
  et~al.}{2018}]{dipierro18b}
{Dipierro} G.,  et~al., 2018, \mn@doi [\mnras] {10.1093/mnras/sty181}, \href
  {https://ui.adsabs.harvard.edu/abs/2018MNRAS.475.5296D} {475, 5296}

\bibitem[\protect\citeauthoryear{{Dong}, {Li}, {Chiang}  \& {Li}}{{Dong}
  et~al.}{2017}]{dong17}
{Dong} R.,  {Li} S.,  {Chiang} E.,   {Li} H.,  2017, \mn@doi [\apj]
  {10.3847/1538-4357/aa72f2}, \href
  {https://ui.adsabs.harvard.edu/abs/2017ApJ...843..127D} {843, 127}

\bibitem[\protect\citeauthoryear{{Dr{\k{a}}{\.z}kowska}, {Li}, {Birnstiel},
  {Stammler}  \& {Li}}{{Dr{\k{a}}{\.z}kowska} et~al.}{2019}]{drazkowska19}
{Dr{\k{a}}{\.z}kowska} J.,  {Li} S.,  {Birnstiel} T.,  {Stammler} S.~M.,   {Li}
  H.,  2019, \mn@doi [\apj] {10.3847/1538-4357/ab46b7}, \href
  {https://ui.adsabs.harvard.edu/abs/2019ApJ...885...91D} {885, 91}

\bibitem[\protect\citeauthoryear{{Fedele}, {Toci}, {Maud}  \&
  {Lodato}}{{Fedele} et~al.}{2021}]{fedele21}
{Fedele} D.,  {Toci} C.,  {Maud} L.,   {Lodato} G.,  2021, \mn@doi [\aap]
  {10.1051/0004-6361/202141278}, \href
  {https://ui.adsabs.harvard.edu/abs/2021A&A...651A..90F} {651, A90}

\bibitem[\protect\citeauthoryear{{Flock}, {Nelson}, {Turner}, {Bertrang},
  {Carrasco-Gonz{\'a}lez}, {Henning}, {Lyra}  \& {Teague}}{{Flock}
  et~al.}{2017}]{flock17}
{Flock} M.,  {Nelson} R.~P.,  {Turner} N.~J.,  {Bertrang} G. H.~M.,
  {Carrasco-Gonz{\'a}lez} C.,  {Henning} T.,  {Lyra} W.,   {Teague} R.,  2017,
  \mn@doi [\apj] {10.3847/1538-4357/aa943f}, \href
  {https://ui.adsabs.harvard.edu/abs/2017ApJ...850..131F} {850, 131}

\bibitem[\protect\citeauthoryear{{Goldreich} \& {Tremaine}}{{Goldreich} \&
  {Tremaine}}{1979}]{goldreich79}
{Goldreich} P.,  {Tremaine} S.,  1979, \mn@doi [\apj] {10.1086/157448}, \href
  {https://ui.adsabs.harvard.edu/abs/1979ApJ...233..857G} {233, 857}

\bibitem[\protect\citeauthoryear{{Goldreich} \& {Tremaine}}{{Goldreich} \&
  {Tremaine}}{1980}]{goldreich80}
{Goldreich} P.,  {Tremaine} S.,  1980, \mn@doi [\apj] {10.1086/158356}, \href
  {https://ui.adsabs.harvard.edu/abs/1980ApJ...241..425G} {241, 425}

\bibitem[\protect\citeauthoryear{Gonzalez, Laibe, Maddison, Pinte  \&
  Ménard}{Gonzalez et~al.}{2015}]{gonzalez15}
Gonzalez J.-F.,  Laibe G.,  Maddison S.~T.,  Pinte C.,   Ménard F.,  2015,
  \mn@doi [Monthly Notices of the Royal Astronomical Society: Letters]
  {10.1093/mnrasl/slv120}, 454, L36

\bibitem[\protect\citeauthoryear{{Gonzalez}, {Laibe}  \& {Maddison}}{{Gonzalez}
  et~al.}{2018}]{gonzalez18}
{Gonzalez} J.~F.,  {Laibe} G.,   {Maddison} S.~T.,  2018, in {Di Matteo} P.,
  {Billebaud} F.,  {Herpin} F.,  {Lagarde} N.,  {Marquette} J.~B.,  {Robin} A.,
    {Venot} O.,  eds, SF2A-2018: Proceedings of the Annual meeting of the
  French Society of Astronomy and Astrophysics. p.~Di

\bibitem[\protect\citeauthoryear{{Goodman} \& {Rafikov}}{{Goodman} \&
  {Rafikov}}{2001}]{2001ApJ...552..793G}
{Goodman} J.,  {Rafikov} R.~R.,  2001, \mn@doi [\apj] {10.1086/320572}, \href
  {https://ui.adsabs.harvard.edu/abs/2001ApJ...552..793G} {552, 793}

\bibitem[\protect\citeauthoryear{{Hsieh} \& {Lin}}{{Hsieh} \&
  {Lin}}{2020}]{hsieh20}
{Hsieh} H.-F.,  {Lin} M.-K.,  2020, \mn@doi [\mnras] {10.1093/mnras/staa2115},
  \href {https://ui.adsabs.harvard.edu/abs/2020MNRAS.497.2425H} {497, 2425}

\bibitem[\protect\citeauthoryear{{Hutchison}, {Price}, {Laibe}  \&
  {Maddison}}{{Hutchison} et~al.}{2016}]{hutchison16}
{Hutchison} M.~A.,  {Price} D.~J.,  {Laibe} G.,   {Maddison} S.~T.,  2016,
  \mn@doi [\mnras] {10.1093/mnras/stw1126}, \href
  {https://ui.adsabs.harvard.edu/abs/2016MNRAS.461..742H} {461, 742}

\bibitem[\protect\citeauthoryear{{Jacquet}, {Balbus}  \& {Latter}}{{Jacquet}
  et~al.}{2011}]{jacquet11}
{Jacquet} E.,  {Balbus} S.,   {Latter} H.,  2011, \mn@doi [\mnras]
  {10.1111/j.1365-2966.2011.18971.x}, \href
  {https://ui.adsabs.harvard.edu/abs/2011MNRAS.415.3591J} {415, 3591}

\bibitem[\protect\citeauthoryear{{Johansen}, {Youdin}  \& {Mac Low}}{{Johansen}
  et~al.}{2009}]{johansen09}
{Johansen} A.,  {Youdin} A.,   {Mac Low} M.-M.,  2009, \mn@doi [\apjl]
  {10.1088/0004-637X/704/2/L75}, \href
  {https://ui.adsabs.harvard.edu/abs/2009ApJ...704L..75J} {704, L75}

\bibitem[\protect\citeauthoryear{{Kanagawa}, {Ueda}, {Muto}  \&
  {Okuzumi}}{{Kanagawa} et~al.}{2017}]{kanagawa17}
{Kanagawa} K.~D.,  {Ueda} T.,  {Muto} T.,   {Okuzumi} S.,  2017, \mn@doi [\apj]
  {10.3847/1538-4357/aa7ca1}, \href
  {https://ui.adsabs.harvard.edu/abs/2017ApJ...844..142K} {844, 142}

\bibitem[\protect\citeauthoryear{{Kley} \& {Crida}}{{Kley} \&
  {Crida}}{2008}]{2008A&A...487L...9K}
{Kley} W.,  {Crida} A.,  2008, \mn@doi [\aap] {10.1051/0004-6361:200810033},
  \href {https://ui.adsabs.harvard.edu/abs/2008A&A...487L...9K} {487, L9}

\bibitem[\protect\citeauthoryear{{Korycansky} \& {Papaloizou}}{{Korycansky} \&
  {Papaloizou}}{1996}]{1996ApJS..105..181K}
{Korycansky} D.~G.,  {Papaloizou} J.~C.~B.,  1996, \mn@doi [\apjs]
  {10.1086/192311}, \href
  {https://ui.adsabs.harvard.edu/abs/1996ApJS..105..181K} {105, 181}

\bibitem[\protect\citeauthoryear{{Laibe} \& {Price}}{{Laibe} \&
  {Price}}{2012}]{laibe12}
{Laibe} G.,  {Price} D.~J.,  2012, \mn@doi [\mnras]
  {10.1111/j.1365-2966.2011.20202.x}, \href
  {https://ui.adsabs.harvard.edu/abs/2012MNRAS.420.2345L} {420, 2345}

\bibitem[\protect\citeauthoryear{{Laibe} \& {Price}}{{Laibe} \&
  {Price}}{2014}]{laibe14}
{Laibe} G.,  {Price} D.~J.,  2014, \mn@doi [\mnras] {10.1093/mnras/stu355},
  \href {https://ui.adsabs.harvard.edu/abs/2014MNRAS.440.2136L} {440, 2136}

\bibitem[\protect\citeauthoryear{{Lega}, {Crida}, {Bitsch}  \&
  {Morbidelli}}{{Lega} et~al.}{2014}]{lega2014}
{Lega} E.,  {Crida} A.,  {Bitsch} B.,   {Morbidelli} A.,  2014, \mn@doi
  [\mnras] {10.1093/mnras/stu304}, \href
  {https://ui.adsabs.harvard.edu/abs/2014MNRAS.440..683L} {440, 683}

\bibitem[\protect\citeauthoryear{{Lin} \& {Youdin}}{{Lin} \&
  {Youdin}}{2017}]{lin17}
{Lin} M.-K.,  {Youdin} A.~N.,  2017, \mn@doi [\apj] {10.3847/1538-4357/aa92cd},
  \href {https://ui.adsabs.harvard.edu/abs/2017ApJ...849..129L} {849, 129}

\bibitem[\protect\citeauthoryear{{Long} et~al.,}{{Long} et~al.}{2018}]{long18}
{Long} F.,  et~al., 2018, \mn@doi [\apj] {10.3847/1538-4357/aae8e1}, \href
  {https://ui.adsabs.harvard.edu/abs/2018ApJ...869...17L} {869, 17}

\bibitem[\protect\citeauthoryear{{Lovascio} \& {Paardekooper}}{{Lovascio} \&
  {Paardekooper}}{2019}]{lovascio19}
{Lovascio} F.,  {Paardekooper} S.-J.,  2019, \mn@doi [\mnras]
  {10.1093/mnras/stz2035}, \href
  {https://ui.adsabs.harvard.edu/abs/2019MNRAS.488.5290L} {488, 5290}

\bibitem[\protect\citeauthoryear{{Masset}}{{Masset}}{2008}]{masset08}
{Masset} F.~S.,  2008, in {Sun} Y.-S.,  {Ferraz-Mello} S.,   {Zhou} J.-L.,
  eds,  Vol. 249, Exoplanets: Detection, Formation and Dynamics. pp 331--346,
  \mn@doi{10.1017/S1743921308016797}

\bibitem[\protect\citeauthoryear{{Masset}}{{Masset}}{2017}]{2017MNRAS.472.4204M}
{Masset} F.~S.,  2017, \mn@doi [\mnras] {10.1093/mnras/stx2271}, \href
  {https://ui.adsabs.harvard.edu/abs/2017MNRAS.472.4204M} {472, 4204}

\bibitem[\protect\citeauthoryear{{Masset} \& {Papaloizou}}{{Masset} \&
  {Papaloizou}}{2003}]{masset2003}
{Masset} F.~S.,  {Papaloizou} J.~C.~B.,  2003, \mn@doi [\apj] {10.1086/373892},
  \href {https://ui.adsabs.harvard.edu/abs/2003ApJ...588..494M} {588, 494}

\bibitem[\protect\citeauthoryear{{McNally}, {Nelson}, {Paardekooper}, {Gressel}
   \& {Lyra}}{{McNally} et~al.}{2017}]{mcnally17}
{McNally} C.~P.,  {Nelson} R.~P.,  {Paardekooper} S.-J.,  {Gressel} O.,
  {Lyra} W.,  2017, \mn@doi [\mnras] {10.1093/mnras/stx2136}, \href
  {https://ui.adsabs.harvard.edu/abs/2017MNRAS.472.1565M} {472, 1565}

\bibitem[\protect\citeauthoryear{{McNally}, {Nelson}, {Paardekooper}  \&
  {Ben{\'\i}tez-Llambay}}{{McNally} et~al.}{2019a}]{mcnally19b}
{McNally} C.~P.,  {Nelson} R.~P.,  {Paardekooper} S.-J.,
  {Ben{\'\i}tez-Llambay} P.,  2019a, \mn@doi [\mnras] {10.1093/mnras/stz023},
  \href {https://ui.adsabs.harvard.edu/abs/2019MNRAS.484..728M} {484, 728}

\bibitem[\protect\citeauthoryear{{McNally}, {Nelson}  \&
  {Paardekooper}}{{McNally} et~al.}{2019b}]{mcnally19a}
{McNally} C.~P.,  {Nelson} R.~P.,   {Paardekooper} S.-J.,  2019b, \mn@doi
  [\mnras] {10.1093/mnrasl/slz118}, \href
  {https://ui.adsabs.harvard.edu/abs/2019MNRAS.489L..17M} {489, L17}

\bibitem[\protect\citeauthoryear{Meru, Rosotti, Booth, Nazari  \& Clarke}{Meru
  et~al.}{2018}]{meru18}
Meru F.,  Rosotti G.~P.,  Booth R.~A.,  Nazari P.,   Clarke C.~J.,  2018,
  \mn@doi [\mnras] {10.1093/mnras/sty2847}, 482, 3678

\bibitem[\protect\citeauthoryear{{Meyer}, {Balsara}  \& {Aslam}}{{Meyer}
  et~al.}{2014}]{meyer14}
{Meyer} C.~D.,  {Balsara} D.~S.,   {Aslam} T.~D.,  2014, \mn@doi [Journal of
  Computational Physics] {10.1016/j.jcp.2013.08.021}, \href
  {https://ui.adsabs.harvard.edu/abs/2014JCoPh.257..594M} {257, 594}

\bibitem[\protect\citeauthoryear{{Nakagawa}, {Sekiya}  \& {Hayashi}}{{Nakagawa}
  et~al.}{1986}]{nakagawa86}
{Nakagawa} Y.,  {Sekiya} M.,   {Hayashi} C.,  1986, \mn@doi [\icarus]
  {10.1016/0019-1035(86)90121-1}, \href
  {https://ui.adsabs.harvard.edu/abs/1986Icar...67..375N} {67, 375}

\bibitem[\protect\citeauthoryear{{Nazari}, {Booth}, {Clarke}, {Rosotti},
  {Tazzari}, {Juhasz}  \& {Meru}}{{Nazari} et~al.}{2019}]{nazari20}
{Nazari} P.,  {Booth} R.~A.,  {Clarke} C.~J.,  {Rosotti} G.~P.,  {Tazzari} M.,
  {Juhasz} A.,   {Meru} F.,  2019, \mn@doi [\mnras] {10.1093/mnras/stz836},
  \href {https://ui.adsabs.harvard.edu/abs/2019MNRAS.485.5914N} {485, 5914}

\bibitem[\protect\citeauthoryear{{Nelson} \& {Papaloizou}}{{Nelson} \&
  {Papaloizou}}{2003}]{nelson03}
{Nelson} R.~P.,  {Papaloizou} J. C.~B.,  2003, \mn@doi [\mnras]
  {10.1046/j.1365-8711.2003.06247.x}, \href
  {https://ui.adsabs.harvard.edu/abs/2003MNRAS.339..993N} {339, 993}

\bibitem[\protect\citeauthoryear{{Ogilvie} \& {Lubow}}{{Ogilvie} \&
  {Lubow}}{2002}]{ogilvie02}
{Ogilvie} G.~I.,  {Lubow} S.~H.,  2002, \mn@doi [\mnras]
  {10.1046/j.1365-8711.2002.05148.x}, \href
  {https://ui.adsabs.harvard.edu/abs/2002MNRAS.330..950O} {330, 950}

\bibitem[\protect\citeauthoryear{{Paardekooper}}{{Paardekooper}}{2014}]{paadekooper14}
{Paardekooper} S.~J.,  2014, \mn@doi [\mnras] {10.1093/mnras/stu1542}, \href
  {https://ui.adsabs.harvard.edu/abs/2014MNRAS.444.2031P} {444, 2031}

\bibitem[\protect\citeauthoryear{{Paardekooper} \& {Mellema}}{{Paardekooper} \&
  {Mellema}}{2004}]{paardekooper04}
{Paardekooper} S.~J.,  {Mellema} G.,  2004, \mn@doi [\aap]
  {10.1051/0004-6361:200400053}, \href
  {https://ui.adsabs.harvard.edu/abs/2004A&A...425L...9P} {425, L9}

\bibitem[\protect\citeauthoryear{{Paardekooper} \& {Mellema}}{{Paardekooper} \&
  {Mellema}}{2006}]{2006A&A...459L..17P}
{Paardekooper} S.~J.,  {Mellema} G.,  2006, \mn@doi [\aap]
  {10.1051/0004-6361:20066304}, \href
  {https://ui.adsabs.harvard.edu/abs/2006A&A...459L..17P} {459, L17}

\bibitem[\protect\citeauthoryear{{Paardekooper}, {McNally}  \&
  {Lovascio}}{{Paardekooper} et~al.}{2020}]{paardekooper20}
{Paardekooper} S.-J.,  {McNally} C.~P.,   {Lovascio} F.,  2020, \mn@doi
  [\mnras] {10.1093/mnras/staa3162}, \href
  {https://ui.adsabs.harvard.edu/abs/2020MNRAS.499.4223P} {499, 4223}

\bibitem[\protect\citeauthoryear{{Price} \& {Federrath}}{{Price} \&
  {Federrath}}{2010}]{price10}
{Price} D.~J.,  {Federrath} C.,  2010, \mn@doi [\mnras]
  {10.1111/j.1365-2966.2010.16810.x}, \href
  {https://ui.adsabs.harvard.edu/abs/2010MNRAS.406.1659P} {406, 1659}

\bibitem[\protect\citeauthoryear{{Ragusa}, {Dipierro}, {Lodato}, {Laibe}  \&
  {Price}}{{Ragusa} et~al.}{2017}]{ragusa17}
{Ragusa} E.,  {Dipierro} G.,  {Lodato} G.,  {Laibe} G.,   {Price} D.~J.,  2017,
  \mn@doi [\mnras] {10.1093/mnras/stw2456}, \href
  {https://ui.adsabs.harvard.edu/abs/2017MNRAS.464.1449R} {464, 1449}

\bibitem[\protect\citeauthoryear{{Riols} \& {Lesur}}{{Riols} \&
  {Lesur}}{2018}]{2018A&A...617A.117R}
{Riols} A.,  {Lesur} G.,  2018, \mn@doi [\aap] {10.1051/0004-6361/201833212},
  \href {https://ui.adsabs.harvard.edu/abs/2018A&A...617A.117R} {617, A117}

\bibitem[\protect\citeauthoryear{{Rodenkirch}, {Rometsch}, {Dullemond}, {Weber}
   \& {Kley}}{{Rodenkirch} et~al.}{2021}]{rodenkirch21}
{Rodenkirch} P.~J.,  {Rometsch} T.,  {Dullemond} C.~P.,  {Weber} P.,   {Kley}
  W.,  2021, \mn@doi [\aap] {10.1051/0004-6361/202038484}, \href
  {https://ui.adsabs.harvard.edu/abs/2021A&A...647A.174R} {647, A174}

\bibitem[\protect\citeauthoryear{{Rowther}, {Meru}, {Kennedy}, {Nealon}  \&
  {Pinte}}{{Rowther} et~al.}{2020}]{rowther20}
{Rowther} S.,  {Meru} F.,  {Kennedy} G.~M.,  {Nealon} R.,   {Pinte} C.,  2020,
  \mn@doi [\apjl] {10.3847/2041-8213/abc704}, \href
  {https://ui.adsabs.harvard.edu/abs/2020ApJ...904L..18R} {904, L18}

\bibitem[\protect\citeauthoryear{{Shakura} \& {Sunyaev}}{{Shakura} \&
  {Sunyaev}}{1973}]{shakura73}
{Shakura} N.~I.,  {Sunyaev} R.~A.,  1973, \aap, \href
  {https://ui.adsabs.harvard.edu/abs/1973A&A....24..337S} {500, 33}

\bibitem[\protect\citeauthoryear{{Stone} \& {Norman}}{{Stone} \&
  {Norman}}{1992}]{stone92}
{Stone} J.~M.,  {Norman} M.~L.,  1992, \mn@doi [\apjs] {10.1086/191680}, \href
  {https://ui.adsabs.harvard.edu/abs/1992ApJS...80..753S} {80, 753}

\bibitem[\protect\citeauthoryear{{Terry}, {Hall}, {Longarini}, {Lodato},
  {Toci}, {Veronesi}, {Paneque-Carre{\~n}o}  \& {Pinte}}{{Terry}
  et~al.}{2022}]{terry22}
{Terry} J.~P.,  {Hall} C.,  {Longarini} C.,  {Lodato} G.,  {Toci} C.,
  {Veronesi} B.,  {Paneque-Carre{\~n}o} T.,   {Pinte} C.,  2022, \mn@doi
  [\mnras] {10.1093/mnras/stab3513}, \href
  {https://ui.adsabs.harvard.edu/abs/2022MNRAS.510.1671T} {510, 1671}

\bibitem[\protect\citeauthoryear{{Toci}, {Lodato}, {Christiaens}, {Fedele},
  {Pinte}, {Price}  \& {Testi}}{{Toci} et~al.}{2020}]{toci20}
{Toci} C.,  {Lodato} G.,  {Christiaens} V.,  {Fedele} D.,  {Pinte} C.,  {Price}
  D.~J.,   {Testi} L.,  2020, \mn@doi [\mnras] {10.1093/mnras/staa2933}, \href
  {https://ui.adsabs.harvard.edu/abs/2020MNRAS.499.2015T} {499, 2015}

\bibitem[\protect\citeauthoryear{{Toci}, {Rosotti}, {Lodato}, {Testi}  \&
  {Trapman}}{{Toci} et~al.}{2021}]{toci21}
{Toci} C.,  {Rosotti} G.,  {Lodato} G.,  {Testi} L.,   {Trapman} L.,  2021,
  \mn@doi [\mnras] {10.1093/mnras/stab2112}, \href
  {https://ui.adsabs.harvard.edu/abs/2021MNRAS.507..818T} {507, 818}

\bibitem[\protect\citeauthoryear{{Tsukamoto}, {Machida}  \&
  {Inutsuka}}{{Tsukamoto} et~al.}{2021}]{tsukamoto21}
{Tsukamoto} Y.,  {Machida} M.~N.,   {Inutsuka} S.,  2021, \mn@doi [\apj]
  {10.3847/1538-4357/abf5db}, \href
  {https://ui.adsabs.harvard.edu/abs/2021ApJ...913..148T} {913, 148}

\bibitem[\protect\citeauthoryear{{Ubeira Gabellini} et~al.,}{{Ubeira Gabellini}
  et~al.}{2019}]{gabellini19}
{Ubeira Gabellini} M.~G.,  et~al., 2019, \mn@doi [\mnras]
  {10.1093/mnras/stz1138}, \href
  {https://ui.adsabs.harvard.edu/abs/2019MNRAS.486.4638U} {486, 4638}

\bibitem[\protect\citeauthoryear{{Vericel}, {Gonzalez}, {Price}, {Laibe}  \&
  {Pinte}}{{Vericel} et~al.}{2021}]{vericel21}
{Vericel} A.,  {Gonzalez} J.-F.,  {Price} D.~J.,  {Laibe} G.,   {Pinte} C.,
  2021, \mn@doi [\mnras] {10.1093/mnras/stab2263}, \href
  {https://ui.adsabs.harvard.edu/abs/2021MNRAS.507.2318V} {507, 2318}

\bibitem[\protect\citeauthoryear{{Veronesi} et~al.,}{{Veronesi}
  et~al.}{2020}]{veronesi20}
{Veronesi} B.,  et~al., 2020, \mn@doi [\mnras] {10.1093/mnras/staa1278}, \href
  {https://ui.adsabs.harvard.edu/abs/2020MNRAS.495.1913V} {495, 1913}

\bibitem[\protect\citeauthoryear{{Whipple}}{{Whipple}}{1972}]{whipple72}
{Whipple} F.~L.,  1972, in {Elvius} A.,  ed., From Plasma to Planet. p.~211

\bibitem[\protect\citeauthoryear{{Youdin} \& {Goodman}}{{Youdin} \&
  {Goodman}}{2005}]{youdin05}
{Youdin} A.~N.,  {Goodman} J.,  2005, \mn@doi [\apj] {10.1086/426895}, \href
  {https://ui.adsabs.harvard.edu/abs/2005ApJ...620..459Y} {620, 459}

\bibitem[\protect\citeauthoryear{{Zhang} et~al.,}{{Zhang}
  et~al.}{2018}]{zhang18}
{Zhang} S.,  et~al., 2018, \mn@doi [\apjl] {10.3847/2041-8213/aaf744}, \href
  {https://ui.adsabs.harvard.edu/abs/2018ApJ...869L..47Z} {869, L47}

\bibitem[\protect\citeauthoryear{{de Val-Borro} et~al.,}{{de Val-Borro}
  et~al.}{2006}]{devalborro06}
{de Val-Borro} M.,  et~al., 2006, \mn@doi [\mnras]
  {10.1111/j.1365-2966.2006.10488.x}, \href
  {https://ui.adsabs.harvard.edu/abs/2006MNRAS.370..529D} {370, 529}

\makeatother
\end{thebibliography}



\clearpage 


\bsp	
\label{lastpage}
\end{document}